\documentclass{article}

\usepackage[table,x11names,dvipsnames,svgnames]{xcolor} 

\usepackage{arxiv}
\usepackage{amsthm}
\usepackage[utf8]{inputenc} 
\usepackage[T1]{fontenc}    
\usepackage{hyperref}       
\usepackage{url}            
\usepackage{booktabs}       
\usepackage{amsfonts}       
\usepackage{nicefrac}       
\usepackage{microtype}      
\usepackage{lipsum}		
\usepackage{graphicx}
\usepackage{doi}
\usepackage{multirow}
\usepackage{comment}
\usepackage{subcaption}
\usepackage{contour}
\usepackage[utf8]{inputenc}
\usepackage{rotating}
\usepackage{hyperref}
\usepackage{pgfplots}
\pgfplotsset{compat=1.17}
\usepackage{physics}
\usepackage{amsmath,amssymb,amsfonts}
\usepackage{algorithmic}
\usepackage{array}
\usepackage{caption}
\usepackage{subcaption}
\usepackage{textcomp}
\usepackage{pgfplots}
\pgfplotsset{compat=newest}
\usepackage{tikz}
\usetikzlibrary{calc}
\usetikzlibrary{arrows}
\usetikzlibrary{patterns}
\usepackage{bondgraphs}
\usepackage[american]{circuitikz}
\usepackage{graphicx}
\usepackage{soul}
\usepackage{doi}
\usepackage{multirow}
\usepackage{comment}
\usepackage{subcaption}
\usepackage{contour}
\usepackage{rotating}
\usepackage{pgfplots}
\usepackage{adjustbox}
\usepackage{makecell}
\usepackage{nicematrix}
\usepackage{float}
\usetikzlibrary{intersections}
\usetikzlibrary{tikzmark}
\usetikzlibrary{shapes,positioning, fit, backgrounds}
\usetikzlibrary{arrows,shapes,   calc, patterns,decorations.pathmorphing,decorations.markings}

\tikzstyle{bag} = [align=center]
\def\BibTeX{{\rm B\kern-.05em{\sc i\kern-.025em b}\kern-.08em
    T\kern-.1667em\lower.7ex\hbox{E}\kern-.125emX}}
\definecolor{agentcolor}{rgb}{0, 0.4627, 0.7608}
\tikzset{
block/.style = {draw, fill=white, rectangle, minimum height=2em, minimum width=2em},
tmp/.style  = {coordinate}, 
sum/.style= {draw, fill=white, circle, node distance=1cm},
input/.style = {coordinate},
output/.style= {coordinate},
container/.style = {draw, agentcolor, fill=none},
pinstyle/.style = {pin edge={to-,thin,black}
}
}
\tikzstyle{bag} = [align=center]
\def\BibTeX{{\rm B\kern-.05em{\sc i\kern-.025em b}\kern-.08em
    T\kern-.1667em\lower.7ex\hbox{E}\kern-.125emX}}
\tikzstyle{densely dashed}=          [dash pattern=on 4pt off 3pt]

\theoremstyle{definition}

\definecolor{agentcolor}{rgb}{0, 0.4627, 0.7608}
\definecolor{mygreen}{RGB}{34, 139, 34}
\definecolor{myred}{RGB}{200, 0, 0}
\definecolor{myblue}{RGB}{0, 0, 153}
\definecolor{MankiwYellow}{RGB}{255, 210, 124}
\definecolor{MankiwGreen}{RGB}{157, 210, 156}
\definecolor{MankiwBlue}{RGB}{188, 220, 245}

\title{Modeling Economic Systems as Multiport Networks}

\author{{ Coen~Hutters} \\
	Delft Center for Systems and Control\\
	Technische Universiteit Delft\\
	Delft, Netherlands \\
	\texttt{C.Hutters-1@tudelft.nl} \\
	\And
	{Max B.~Mendel} \\
	Delft Center for Systems and Control\\
	Technische Universiteit Delft\\
	Delft, Netherlands \\
	\texttt{M.B.Mendel@tudelft.nl} \\
}

\renewcommand{\shorttitle}{\textit{arXiv} Template}

\hypersetup{
pdftitle={A template for the arxiv style},
pdfsubject={q-bio.NC, q-bio.QM},
pdfauthor={David S.~Hippocampus, Elias D.~Striatum},
pdfkeywords={First keyword, Second keyword, More},
}

\begin{document}
\maketitle

\begin{abstract}
In this paper, we demonstrate how multiport network theory can be used as a powerful modeling tool in economics. 
The critical insight is using the port concept to pair the flow of goods (the electrical current) with the agent’s incentive (the voltage) in an economic interaction. 
By building networks of agents interacting through ports, we create models with multiple levels of abstraction; from the macro level down to the micro level.  
We are thereby able to model complex macroeconomic systems whose dynamical behavior is emergent from the micro level.
Using the LTSpice circuit simulator, we then design and analyze a series of example systems that range in complexity from the textbook Robinson Crusoe economy to a model of an entire economy.
\end{abstract}
\newcommand{\MPC}{\text{MPC}}
\newcommand{\MPL}{\text{MPL}}

\newcommand{\Y}{\boldsymbol{y}}
\newcommand{\Z}{\boldsymbol{z}}
\newcommand{\HH}{\boldsymbol{h}}
\newcommand{\G}{\boldsymbol{g}}
\newcommand{\T}{\boldsymbol{t}}
\newcommand{\E}{\boldsymbol{e}}

\section{Introduction}
\label{sec:introduction}
Multiport network theory has proven to be a powerful tool in electronics and mechatronics, especially when this concerns the design and analysis of highly complex systems.  The theory is based on electrical circuit theory, to which it adds several key features.  These include, in particular, the ability to represent the system at different levels of abstraction, its modular nature, and its high level of computational efficiency \cite{breedveld2004port, Karnopp2012, alexander2012fundamentals}.

In this paper, we leverage these features to model, design, and analyze economic systems. 

It builds on a previous paper \cite{Hutters2024} of ours, in which we develop {economic circuit theory}.  Economic circuits are electronic circuits whose variables and parameters are replaced by their economic analogs.  
Their primary advantage is that circuits are inherently dynamic, providing explicit time differential equations for the flow of goods and their price \cite{Mendel2023}. 
A critical disadvantage of circuit theory that we have found is that they are inefficient for modeling complex systems.  Economic circuits are built by interconnecting elementary agents or components into networks. As such, they are built exclusively at the micro level and, therefore, rapidly become unmanageable when they grow in size.    

Using multiport theory, models can be designed at any level of abstraction.    It provides for modules called multiports or $n$-ports,  which internally contain entire circuits but are treated like ``black boxes.''  A multiport can represent any economic unit or agent that is appropriate to a particular level of abstraction: from the macro level of an entire economy, down through sectors, subsectors, individuals, to the micro level of components.   The theory thus fits in a more general trend in economics to account for micro-level heterogeneity at the macro level \cite{Stiglitz2018, romer2016trouble, Fagiolo2007,platt2020comparison}.

By modeling in a top-down fashion, going from the macro level through intermediate levels down to the micro level, the design of complex systems becomes manageable.  In the economic tradition, 
 the Macroeconomic Dynamic Stochastic General Equilibrium (DSGE) models \cite{smetswouters2004, delnegro2013frbny, Christiano2018} are built in this fashion.  It is assumed that macro-level sectors can be represented by what are known as representative agents.  
 
 The concept of representative agents has come under criticism in the economic literature for failing to account for heterogeneity at lower levels of abstraction
 \cite{Stigliz2017, Colander2008, Farmer2009, Dilaver2018}.   However, by treating a representative agent as a multiport, a fully dynamic description of its micro-level behavior at the component level is obtained.  Moreover, the heterogeneity of representative agents can be specified at various levels of abstraction between the micro and macro levels.  We demonstrate this in Section \ref{sec:Applications}, where we compare a DSGE model from the literature to a multiport implementation.



 Inversely, by working in a bottom-up fashion, the behavior at the macro level is treated as emergent from its specification as a circuit at the micro level.  This is comparable to how Agent-Based Computational Economics (ACE)  models are built \cite{LeBaron2008, farmer2009virtues, Lebaron2016, Westphal2020}. 
 
These, however, have been criticized for lacking a formalism for aggregating agents, instead relying on ad-hoc methods  \cite{DALY2022}.     In addition, the models are notoriously computationally inefficient: Their reliance on Monte-Carlo simulations renders them relatively computationally intensive, difficult to calibrate, replicate, or extend systematically \cite{Fagiolo2017, mcculloch2022}.  

In contrast, multiport theory provides a structured formalism for aggregating micro-level agents into higher-level economic units.  We detail this in Section \ref{sec:AgentsAsMultiports}.   In addition, a multiport network is readily implemented on modern circuit simulators like LTSpice, which are known to be computationally efficient and are capable of handling complex networks.  This is demonstrated in Section \ref{sec:Applications} for a wide variety of applications.   


The paper is organized as follows:  First, in Section \ref{sec:EconomicMOdelsasMultiportNetworks}, 
we outline the economic analogs to the basic concepts of the theory, like ports, interconnections, and the multiports themselves.  Then, in Section \ref{sec:AgentsAsMultiports}, we examine the multiport agent, how its endogenous dynamic behavior is modeled by an internal circuit, and how this is summarized by a behavioral model.   In Section \ref{sec:InteractionsInterconnections}, we show how the different multiport interconnection patterns allow us to aggregate agents to higher levels of abstraction.   Finally, in \ref{sec:Applications}, we provide several example applications that we implemented on the LTSpice circuit simulator to demonstrate how the theory can be used to design and analyze dynamic economic systems.

\section{Economic Models as Multiport Networks}
\label{sec:EconomicMOdelsasMultiportNetworks}
\begin{figure*}[h]
\centering
\begin{subfigure}{0.49\textwidth}
    \centering
\includegraphics[width=\textwidth]{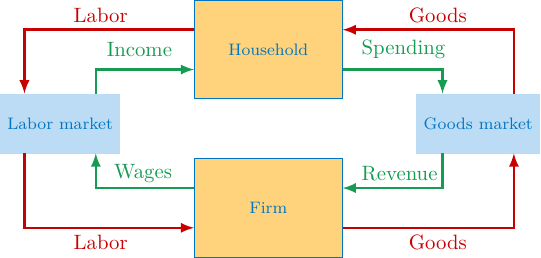}
        \caption{Textbook circular flow model (adapted from \cite{Mankiw2016}).}
        \label{fig:circflow}
\end{subfigure}
\hfill
\begin{subfigure}{0.5\textwidth}
    \centering   
\includegraphics[width=\textwidth]{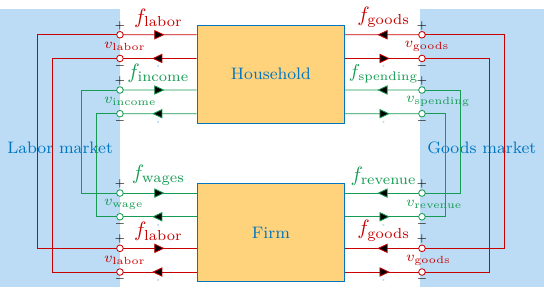}
    \caption{Circular flow model in Figure \ref{fig:circflow} as a multiport network.}
    \label{fig:multiportflow}
\end{subfigure}
\begin{subfigure}{.49\textwidth}
    \centering
    \includegraphics[width=\textwidth]{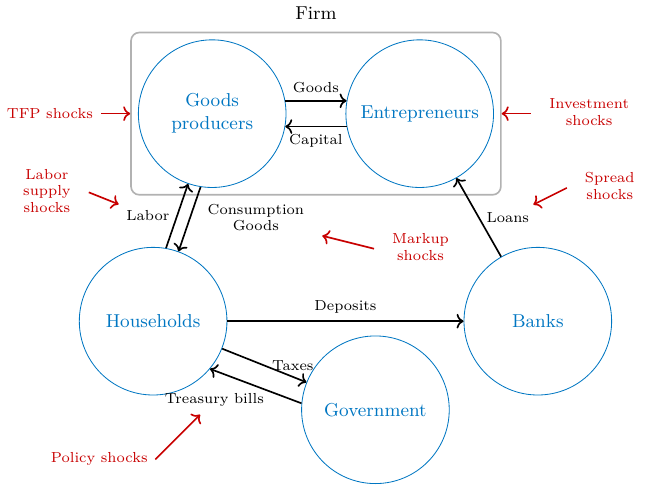}
    \caption{Stylized description of a DSGE model (adapted from \cite{frbny_dsge_model}).}
    \label{fig:DSGE}
\end{subfigure}
\begin{subfigure}{.49\textwidth}
    \centering
    \includegraphics[width=.95\textwidth]{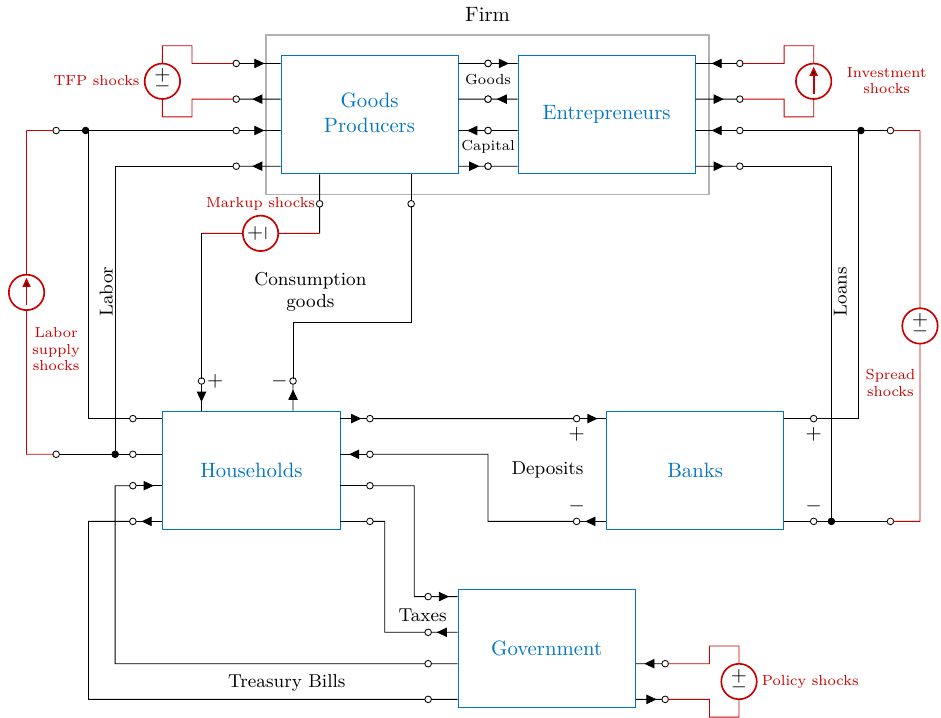}
    \caption{The DSGE model in Figure \ref{fig:DSGE} as a multiport network.}
    \label{fig:DSGEmulti}
\end{subfigure}
    \caption{
    Traditional economic models (left) and their multiport adaptations (right).
  }
    \label{}
\end{figure*}

In the economic literature, authors draw schematic representations of their models to emphasize the logical structure.  Figures \ref{fig:circflow} and \ref{fig:DSGE} are typical examples, the former of a circular flow model, as given in a popular macro text \cite{Mankiw2016}, and the latter of a DSGE model by the New York Federal Reserve Bank \cite{frbny_dsge_model}.  The blocks or bubbles represent economic agents, which could be individuals, companies, or entire sectors of the economy.  The arrows connecting them represent the direction of the flow of either money, goods, or services.  

Engineers draw similar representations when designing multiport models.  In Figures~\ref{fig:multiportflow} and \ref{fig:DSGEmulti}, we give the multiport networks corresponding to these models.  The agents correspond to specific multiports, which are similarly pictured as boxes.  The connections, however, appear as loops rather than arrows.   They enter a multiport at a pair of terminals rather than a single point of entry.   Such a pair is known as a port.  A port allows an agent to communicate the incentive it requires for supplying the flow, together with the flow itself.   This is further detailed in Section \ref{sec:InteractionsWithPorts}.   Table \ref{tab: Signals} reproduces the relationship between the electrical and economic terminology from \cite{Hutters2024}.




This is an important departure from the economic models in Figures \ref{fig:circflow} and \ref{fig:DSGE}. These show what flows where, but not why.  With conjugate incentives at its ports, an agent's endogenous dynamics ---as specified by its internal economic circuit--- determine the economic forces that cause the flows. 

Although in general markets can be viewed as sophisticated multiport agents in their own right, we have modeled them in Figure \ref{fig:multiportflow} as flow-through entities (blue).  They pair two associated flows, of which one is a good or service, and the other is the money provided as compensation:  payment for the goods and remuneration for the labor.  This makes a compelling connection with business practice. For instance, the revenue and goods ports at the firm can, together, be thought of as a point of sale (POS).  We develop this connection further in Section \ref{sec:InteractionsAndTransactios}.

\begin{table}[h!]
    \centering
    \caption{Analogy between economic and electrical signals \cite{Hutters2024}.}
    \label{tab: Signals}
    \begin{tabular}{lllc}
        \toprule
         \textbf{Economic}& \textbf{Electrical} & \textbf{Relation} & \textbf{Units}\\ 
        \midrule
        Flow &Current &$f=\dv[]{q}{t}$&$\frac{\#}{\text{yr}}$\\
         Incentive & Voltage &$v=\dv[]{p}{t}$&$\frac{\$}{\#\text{yr}}$\\
          
        \midrule 
        Stock  &Electric Charge & $q=\int f\dd t$&$\#$\\
        Reservation Price & Magnetic Flux & $ p=\int v \dd t$&$\frac{\$}{\#}$\\
        \midrule
        Surplus Allocation Rate & Power &  $P = fv$ & $\frac{\$}{\text{yr}^2}$ \\
        \bottomrule
    \end{tabular}
\end{table}

\section{Agents as multiports}
\label{sec:AgentsAsMultiports}

\subsection{Communication with Signals}
In the circular flow model of Figure \ref{fig:circflow}, a flow is given by a single number specifying its equilibrium value.  In the multiport network model of Figure \ref{fig:multiportflow}, a flow is, instead, given by a signal specifying the flow's behavior as a function $f = f(t)$ of time.   

In a multiport network, the interaction between agents at their ports also includes an incentive $v$.
Similar to how $f$ can be seen as a rate of change in goods, $v$ can be seen as a rate of change in price \cite{Mendel2023} and is therefore also regarded as a signal, i.e., a function $v = v(t)$. 

Agents communicate the quantities they demand or the incentives they require by sending entire signals.  A single equilibrium value constitutes the special case of a constant or DC signal.  Although in general these can take on a wide variety of shapes, two fundamental types of signals are of particular interest in economics, as they are in engineering.  

First are sinusoidal or AC signals.  These appear in cyclical behavior which manifests itself in business cycles \cite{christiano1998business}, credit cycles \cite{koopman2009credit}, Kondratieff waves \cite{metz2011kondratieff}, and other seasonalities \cite{granger1978seasonality}. Second are 
exponentially decaying or growing signals.  These correspond to exponential discounting or exponential growth \cite{Mendel2023}.  Arbitrary signals can be constructed by superimposing seasonal, exponential, and stationary signals.  

Like their engineering counterparts, the combination of cyclical and discounting signals is more conveniently represented in the frequency domain. Laplace transforming $f$ and $v$ gives, respectively, signals $F$ and $V$ which are functions $F= F(s)$ and $V = V(s)$ of the complex frequency $s = \sigma + i \omega$.  For economics, we think of $\sigma$ as the discount rate (see, e.g., \cite{Varian2010}) and $\omega$ as the frequency of the cyclical component of the signal.  A particular economic cycle or discount rate can then be represented by a single value in the spectrum.  

\subsection{Interactions and Transactions}
\label{sec:InteractionsAndTransactios}
An interaction between agents involves a conjugate pair  $(f,v)$ of signals, where both the flow $f$ and the corresponding incentive $v$ are communicated.  This is a significant departure from economic models, where only the flow is communicated.   It is the presence of a conjugate incentive that allows us to accurately represent the dynamic behavior as emergent from the micro level.   

This definition of an interaction also allows us to keep track of the flow of economic surplus to or from an agent, i.e., the rate at which the surplus is allocated in time to the agent.  This corresponds to the power in an electrical network (see further \cite{Mendel2023}  and also \cite{Hutters2024}).  It is the product of the flow of goods with the incentive:
\begin{equation}
    P = v f
    \label{eq:Power}
\end{equation} 

A transaction involves two interactions:   one for the goods or services, and one for the accompanying money.  In Figure \ref{fig:multiportflow}, we indicate two types of transactions: one for the purchase and sale of the goods and another for the hiring and firing of labor.  Whereas interactions involve only an exchange of either a good or a service, in a transaction, this is paired with the exchange of money.   

The definition of a transaction allows us to make the connection with accounting and business practice.  Two quantities are important:
\begin{itemize}
    \item \textit{Transaction Price: } Ratio of the flow of money to the flow of the good or service.  For example, in Figure \ref{fig:multiportflow} the ratio $f_\text{spending}/f_\text{goods}$ gives the current purchase price, in \$ per item, in the goods market.  
    \item \textit{Accounting Profit:} Difference between incoming and outgoing flow of money from two markets.   For example, in Figure \ref{fig:multiportflow} the difference $f_\text{revenue} - f_\text{wages}$ gives the rate of profit, in \$ per time,  for the firm.  
\end{itemize}
The accounting notions of price and profit are distinct from the economic notions of the reservation price and surplus allocation rate, as they are defined in Table \ref{tab: Signals}.  The latter concepts are endogenous to the agent, and a notion of a transaction is not required for their definition.  

\subsection{Interactions with Ports}
\label{sec:InteractionsWithPorts}
In circular flow or DSGE models, the arrows giving the inter-agent interactions are drawn in a rather informal manner.   
In multiport networks, in contrast, these are formalized through the notion of a port.  When designing a network, we first enumerate the distinct goods, services, or types of money with which an agent interacts.  Then, we open a port for each of them.  An agent exchanging $n$ types of goods, services, or types of money is thus represented by an $n$-port.

Schematically, ports consist of two terminals, which are indicated by a pair of white bullets in Figure \ref{fig:multiportflow}. In the economic interpretation of these terminals, the $+$ and $-$  signs give the polarity of the incentive \cite{Hutters2024}, indicating whether it is a proper incentive (positive) or a desire (negative).  The arrows specify the positive direction of the flow and allow us to distinguish between receipt and delivery. In contrast to the circular flow model, there is no implied direction for the goods or money, and, like currents in circuit theory, they can be either positive or negative.   

A valid port is one to which the port condition applies \cite{PSpiceCircuit2007circuit}. The condition dictates that the incoming flow at a terminal equals the outgoing flow at the other terminal. In electrical networks, it ensures that the network does not violate Kirchhoff's current law.  In economic applications, it serves to impose the physical clearing of the goods within the network (see \cite{Hutters2024}). 



\subsection{Exogenous-Endogenous Shocks with Independent-Dependent Signals}
In the economic literature, a distinction is made between exogenous variables, which originate from outside of the model, and endogenous variables, which are determined by the model \cite{mankiwMacroeconomics2013}.   In a typical analysis, exogenous shocks are imposed, and then the response of endogenous variables of interest is investigated.  

In multiport modeling,  we generalize this concept and 
apply the distinction to each agent individually, at any level of abstraction.  Multiports have a well-defined system boundary. Exogenous signals are, thus, necessarily imposed at a port.   In multiport theory, such signals are known as independent signals.  The endogenously determined response is referred to as a dependent signal.

There is a limit on the number of independent signals which can be exogenously imposed on a multiport.  For an arbitrary $n$-port,  there are $2n$ signals consisting of an incentive and a flow for each port.  At a maximum, $n$ of these can be chosen to be independent.  It does not matter which ports these signals belong to. The remaining $n$ signals are then necessarily dependent and form the endogenously generated response of the agent.  If there are fewer than $n$ exogenously imposed signals, one has the freedom to treat some signals as if they were independent.

Using this distinction at a particular level of abstraction allows us to analyze networks in a causal manner.   Starting off with an exogenous shock to a particular agent, we can causally track the response by taking its dependent signal and treating it as an independent signal of a counterparty agent.   This process then repeats until we arrive at the agent whose dependent variable forms the response of interest.    Any freedom to choose which variables are dependent can be exploited to investigate different causal chains leading to the response.   Section \ref{sec:Applications} provides several demonstrations of this technique.


\subsection{Agent Dynamic Behavior with Parameter Models}


To determine the response of an agent to exogenously imposed signals, we must establish the relationship between the dependent and independent signals.  Formally, such a relationship is an operator between function spaces.  We refer to it as the agent's dynamic behavioral operator.  

In this paper, we limit ourselves to linear dynamics and, hence, the behavioral operator becomes a linear operator.   For these, a frequency domain representation is the most convenient, because the operator's action can be described using linear algebra as follows:  For a general $n$-port, we collect the Laplace transforms of  $n$ independent signals into a vector $\boldsymbol{U}$ and the $n$ dependent ones into a vector $\boldsymbol{Y}$.  The relationship is then given by 
\begin{equation}
    \boldsymbol{Y} = \E \,  \boldsymbol{U}
    \label{eq:GeneralElasticity}
\end{equation}
where $\E$ is a $n \times n$ complex matrix whose entries are also functions of $s$. In multiport theory, these are commonly referred to as parameters.   The behavioral operator is also referred to as a parameter model.   For applications to economics, we can think of the parameters as elasticities and the matrix $\E$ as giving the $\boldsymbol{U}$ elasticity of $\boldsymbol{Y}$, although other interpretations are often relevant (see Section \ref{sec:AgentsAsEquivalent2Ports}).   

In multiport theory, a convenient classification of parameter models exists for 2-ports (see, e.g. \cite{alexander2012fundamentals}).  In this case, the behavioral operator is represented by a $2 \times 2$ complex matrix.  For a 2-port, there are four distinct choices of independent variables, which lead to the four models shown 
in Table \ref{tab:TwoPort}.  In the following subsections, we interpret these from an economic perspective with the aid of the supply chain intermediary in Figure \ref{fig:MatchingNetwork} of the example in Section \ref{sec:bullwhip}.  In addition, we show how the dynamic behavioral operator emerges from the dynamics at the circuit level.  We refer the reader to \cite{Hutters2024} for how this is done.   


\begin{table}[ht]
    \centering
    \caption{Economic elements and their electrical analogs with their behavior in the frequency domain (adapted from \cite{Hutters2024}). $\varepsilon,$ $k$, and $b$ are the elasticities of demand, storage capacity, and economic friction, respectively. }
    \label{tab:elements}
    \begin{tabular}{llll}
        \toprule
        \textbf{Economic} & \textbf{Electrical} & \textbf{Behavior}& \textbf{Symbol} \\
        \midrule
        
        Demand & Inductor & \(V = \frac{1}{{\varepsilon}}{s}F\) & 
        \begin{circuitikz}[scale=0.6, transform shape]
            \draw (0,0) to[L] (2,0);
        \end{circuitikz} \\
        
        Storage & Capacitor & \( V = \frac{1}{sk}F\) & 
        \begin{circuitikz}[scale=0.6, transform shape]
            \draw (0,0) to[C] (2,0);
        \end{circuitikz} \\[5pt]
        
        Friction & Resistor & \(V = \frac{1}{b} F\) & 
        \begin{circuitikz}[scale=0.6, transform shape]
            \draw (0,0) to[R] (2,0);
        \end{circuitikz} \\
        \bottomrule
    \end{tabular}
\end{table}

\begin{table*}[h]
    \centering
     \caption{Parameter models for a 2-port agent.  The agent acts as a middleman between its upstream and downstream counterparties.   It maintains a price at the inductor, an inventory at the capacitor, and accrues transaction costs at the resistor (see Table \ref{tab:elements}).  Upstream and downstream signals are subscripted with a $u$ and $d$, respectively.
}
    \label{tab:TwoPort}
    \newcolumntype{P}[1]{>{\raggedright\arraybackslash}m{#1}}
    \renewcommand{\arraystretch}{1.2} 
    \begin{tabular}{|l|P{2.2cm}|c|c|}
        \hline
        \textbf{Type}    &\textbf{Description} & \textbf{Circuit}&\textbf{Parameter Model}\\ \hline
        Admittance &  { Flexibility: Agent ``admits'' the upstream supply downstream.}

        &
      
        \begin{circuitikz}[baseline={(current bounding box.center)}]
        \ctikzset{bipoles/length=1cm}
        \newcommand{\h}{1.5cm}
        \newcommand{\w}{1.5*\h}
        \node[draw,agentcolor,,rectangle, minimum width=\w, minimum height=\h] (A) at (0,0) {\footnotesize $ $};

            \draw (A.155) to[short, i<_=$F_u$] ++(-\w/5,0) node[ocirc](A1){};
            \draw (A.205) to[short,] ++(-\w/5,0) node[ocirc](A2){};
            \draw (A.25)  to[short, i<^={\scriptsize $F_d$}] ++(\w/5,0) node[ocirc](A3){};
            \draw (A.335) to[short,] ++(\w/5,0) node[ocirc](A4){};
            \draw (A1) to[open, v=$V_u$] (A2);
            \draw (A3) to[open, v=$V_d$] (A4);
            \draw(A2) to++(-.6cm,0) coordinate(a){};
            \draw (A1) to++(-.6cm,0) to[american voltage source] (a);
            
            \draw (A4) to++(.6cm,0) coordinate(b) {};
            \draw (A3) to++(.6cm,0) to[american voltage source] (b);
           
      \ctikzset{bipoles/length=.7cm}
         \draw  (A.155)  to[L,l_=$\varepsilon$] ++(\w/2,0) coordinate(a1)  to[] (A.25)
         (A.205) to  ++(\w/2,0) coordinate(a2) to (A.335)
         (a1) to[R=$b$,*-] ++(0,-.3*\w) to[C=$k$,-*] (a2)
                ;
         \end{circuitikz}   &  ${
       {\begin{pmatrix}
            F_u\\
            F_d
        \end{pmatrix}}
        =
        \underbrace{\begin{pmatrix}
            \frac{\varepsilon}{s}  & -\frac{\varepsilon}{s} \\
-\frac{\varepsilon}{s} & \frac{s^2+\varepsilon bs+\varepsilon k}{bs^2 + ks}
        \end{pmatrix}}_{\Y = \Z^{-1}}
       
        \begin{pmatrix}
            V_u\\
            V_d
        \end{pmatrix}
        }$   \\  \hline
        Impedance  & 
       Stickiness: Agent ``impedes'' the upstream supply  from reaching downstream   
        &       
        \begin{circuitikz}[baseline={(current bounding box.center)}]
        \ctikzset{bipoles/length=1cm}
        \newcommand{\h}{1.5cm}
        \newcommand{\w}{1.5*\h}
        \node[draw,agentcolor, rectangle, minimum width=\w, minimum height=\h] (A) at (0,0) {};

            \draw (A.155) to[short, i<_=$F_u$] ++(-\w/6,0) node[ocirc](A1){};
            \draw (A.205) to[short,] ++(-\w/6,0) node[ocirc](A2){};
            \draw (A.25)  to[short, i<^=$F_d$] ++(\w/6,0) node[ocirc](A3){};
            \draw (A.335) to[short,] ++(\w/6,0) node[ocirc](A4){};
            \draw (A1) to[open, v=$V_u$] (A2);
            \draw (A3) to[open, v=$V_d$] (A4);
          
            \draw (A1) to++(-.7cm,0) coordinate(a) {};
            
            \draw (A2) to++(-.7cm,0) to[american current source] (a);
            \draw (A3) to++(.7cm,0) coordinate(b) {};
            \draw (A4) to++(.7cm,0) to[american current source] (b);

            \ctikzset{bipoles/length=.7cm}
         \draw  (A.155)  to[L,l_=$\varepsilon$] ++(\w/2,0) coordinate(a1)  to[] (A.25)
         (A.205) to  ++(\w/2,0) coordinate(a2) to (A.335)
         (a1) to[R=$b$,*-] ++(0,-.3*\w) to[C=$k$,-*] (a2)
                ;
    \end{circuitikz} & $
        \begin{pmatrix}
            V_u\\
            V_d
        \end{pmatrix}
        =
        \underbrace{        
        \begin{pmatrix}
           \frac{ s^2+\varepsilon b s+\varepsilon k}{\varepsilon s}
         & \frac{bs+k}{s}\\
           \frac{bs+k}{s} & \frac{bs+k}{s}
        \end{pmatrix}}_{\Z = \Y^{-1}}
        \begin{pmatrix}
            F_u\\
            F_d
        \end{pmatrix}
        $ \\ \hline
         Transmission &  
         Surplus Distribution:  Agent ``transmits'' the upstream surplus downstream
        &
        \begin{circuitikz}[baseline={(current bounding box.center)}]
        \ctikzset{bipoles/length=1cm}
        \newcommand{\h}{1.5cm}
        \newcommand{\w}{1.5*\h}
        \node[draw,agentcolor,rectangle, minimum width=\w, minimum height=\h] (A) at (0,0) {};
            
            \draw (A.155) to[short, i<_=$F_u$] ++(-\w/6,0) node[ocirc](A1){};
            \draw (A.205) to[short,] ++(-\w/6,0) node[ocirc](A2){};
            \draw (A.25)  to[short, i<^=$F_d$] ++(\w/6,0) node[ocirc](A3){};
            \draw (A.335) to[short,] ++(\w/6,0) node[ocirc](A4){};
            \draw (A1) to[open, v=$V_u$] (A2);
            \draw (A3) to[open, v=$V_d$] (A4);
            \draw (A2) to++(-1.4cm,0) coordinate(a) {};
            \draw[white] (A1) to ++(-1.4cm,0) coordinate(b) {};
            \draw (b) to[american voltage source] (a);
            \draw (b) to[american current source] (A1);

            \ctikzset{bipoles/length=.7cm}
         \draw  (A.155)  to[L,l_=$\varepsilon$] ++(\w/2,0) coordinate(a1)  to[] (A.25)
         (A.205) to  ++(\w/2,0) coordinate(a2) to (A.335)
         (a1) to[R=$b$,*-] ++(0,-.3*\w) to[C=$k$,-*] (a2)
                ;
    \end{circuitikz} 
       & $ 
            
             \begin{pmatrix}
            V_d\\ F_d
            \end{pmatrix}
            =
            \underbrace{
            \begin{pmatrix}
            1 & \frac{bs^2+ks}{\varepsilon bs+\varepsilon k}\\
            \frac{s}{bs+k} & \frac{bs+k}{\varepsilon}\\
            \end{pmatrix}
             }_{\T}
            \begin{pmatrix}
            V_u\\ -F_u
            \end{pmatrix}

        $
       \\ \hline
        Hybrid  &  Agent  ``impedes'' the upstream supply and  ``admits'' the downstream supply (and vice versa).
        &
        \begin{circuitikz}[baseline={(current bounding box.center)}]
        \ctikzset{bipoles/length=1cm}
        \newcommand{\h}{1.5cm}
        \newcommand{\w}{1.5*\h}
        \node[draw,agentcolor, rectangle, minimum width=\w, minimum height=\h] (A) at (0,0) {};

            \draw (A.155) to[short, i<_=$F_u$] ++(-\w/6,0) node[ocirc](A1){};
            \draw (A.205) to[short,] ++(-\w/6,0) node[ocirc](A2){};
            \draw (A.25)  to[short, i<^=$F_d$] ++(\w/6,0) node[ocirc](A3){};
            \draw (A.335) to[short,] ++(\w/6,0) node[ocirc](A4){};
            \draw (A1) to[open, v=$V_u$] (A2);
            \draw (A3) to[open, v=$V_d$] (A4);
            \draw (A1) to++(-.7cm,0) coordinate(a) {};
            
            \draw (A2) to++(-.7cm,0) to[american current source] (a);
            \draw (A4) to++(.7cm,0) coordinate(b) {};
            \draw (A3) to++(.7cm,0) to[american voltage source] (b);

            \ctikzset{bipoles/length=.7cm}
         \draw  (A.155)  to[L,l_=$\varepsilon$] ++(\w/2,0) coordinate(a1)  to[] (A.25)
         (A.205) to  ++(\w/2,0) coordinate(a2) to (A.335)
         (a1) to[R=$b$,*-] ++(0,-.3*\w) to[C=$k$,-*] (a2)
                ;
    \end{circuitikz} 
    &$
        \begin{pmatrix}
            V_u\\
            F_d
        \end{pmatrix}
        =
        \underbrace{\begin{pmatrix}
            \frac{bs^2+ks}{\varepsilon bs+\varepsilon k} & 1\\
            -1 & \frac{s}{bs+k}
        \end{pmatrix}}_{\HH=\G^{-1}}
        \begin{pmatrix}
            F_u\\
            V_d
        \end{pmatrix}
        $
        \\ \hline
       
    \end{tabular}
\end{table*}

\subsubsection*{(a) Admittance or $\Y$-parameters}
The first row of Table \ref{tab:TwoPort} contains what are known as admittance or $\Y$ parameters in multiport theory. They loosely correspond to the notion of flexibility in economics.  
 The incentives are imposed, and the agent responds with the flows demanded or supplied.  A specific example is displayed in Figure \ref{fig:Consumer} and discussed in the accompanying text.   

The parameter values follow readily from the circuit.  The circuit acts as an ``incentive divider,''  the economic analog to a voltage divider.  Integrating the net incentive $V_d - V_u$ gives the agent's reservation price, i.e., $p = \frac{1}{s}(V_d - V_u)$.  A subsequent multiplication with the inductor's price elasticity gives the upstream flow, i.e., $F_u = \varepsilon p$.    This completes the first row of the matrix.  
The second row gives the amount $F_d$ that is ``admitted'' downstream.     This consists of the negative of the upstream flow adjusted with the amount coming out of or going into storage (the capacitor).   This latter amount depends on the downstream incentive alone and modifies the bottom right parameter to the given expression.      


\subsubsection*{(b) Impedance $\Z$-parameters}

The impedance or $\Z$ parameter matrix in the second row of Table \ref{tab:TwoPort} is the matrix inverse $\Z = \Y^{-1}$ of the admittance parameters. Stickiness is a general term used in economics as an antonym for flexibility.

The central bank in Figure \ref{fig:bank} is a natural fit for $\Z$-parameters.  It is subjected to two exogenous flows, a target and realized inflation rate, and responds by incentivizing the market to change these.  

The parameters for the example circuit are established by reasoning as follows: The imbalance $ F_u-F_d$ between the exogenous upstream and downstream flows accumulates in the storage into a stock $q = {1}/{s}( F_u-F_d )$ at the capacitor.   Multiplication with the storage elasticity $k$ gives the incentive $kq$ measuring the relative scarcity of the goods.  The premiums required to overcome the economic friction are added to this.  These are proportional to the flows with constant $b$.  For the top left parameter, we also need to take into account the incentive that represents the force of demand for the upstream supply.     


\subsubsection*{(c) Transmission or $\T$-parameters}

The transmission, chain, cascade,  or $\T$ parameters are used to model how the economic surplus can propagate through a network.   They are characterized by having both the independent signals on a single port called the input port.  The pair of dependent signals appears on the remaining output port.  This condition allows one to calculate the rate at which surplus  is delivered at the ports as the convolutions
\begin{equation}
    F_u * V_u \qquad\text{and} \qquad F_d * V_d .
\end{equation}
In the time domain, the convolutions revert to a regular product (\ref{eq:Power}). 

In electrical networks, $\T$ parameters are useful in analyzing power transmission systems.  The analogy between energy and surplus makes them useful for economics when analyzing the allocation of surplus between consumers and producers.  

In particular, as the alternate name ``chain parameters'' might already suggest, they are the appropriate choice when modeling supply chains; in fact, the example circuit was taken from the chain in Figure \ref{fig:MatchingNetwork}.  When one thinks of the chain as being supply-driven, the upstream port should be thought of as the input port and the downstream one as the output.  (For demand-driven chains, the inverse choice is appropriate.)  When chained together with a producer at the upstream end and a consumer at the downstream end, the example explicitly calculates the dynamics of how the surplus is distributed between the two, and how much of it is lost in the chain itself.

\subsubsection*{(d) Hybrid $\HH$ and $\G$-parameters }

Hybrid parameters are used when independent signals consist of the flow on one port and the incentive on the other port. 
For 2-ports, there are two types of parameter models:  $\HH$ parameters are typically used when the incentive is thought of as the cause, and $\G = \HH^{-1}$ parameters when it is the flow.   

In electrical networks, hybrid parameters are useful for amplifiers, transistors, and controlled sources.  When these elements are applied to economics, they are likewise useful, as evidenced by the examples in Figures \ref{fig:Multivarprod} and \ref{fig:dimprod}.

The agent in Table \ref{tab:equivalentmodels} incentivizes its upstream supplier by passing along its customer's incentive $V_d$, together with its own incentives for storing, overcoming friction, and demand.   This gives the dependent $V_u$.   The other dependent signal, $F_d$, is determined in a similar manner.  




\subsection{Micro-Level Dynamical Behavior }

We specify the dynamical behavior of an agent at the micro level, i.e., the lowest level of abstraction.  
This is the level where economic circuit theory applies.  
Economic circuits are the economic analogs to electronic circuits, and a broad range of dynamic economic behavior can be modeled in this manner.
Table \ref{tab:elements} contains the analogy between elementary passive electrical components and economic behavior and the corresponding behavioral law in the frequency domain.
We refer the reader to \cite{Hutters2024} for a comprehensive treatment of the theory in the time domain.

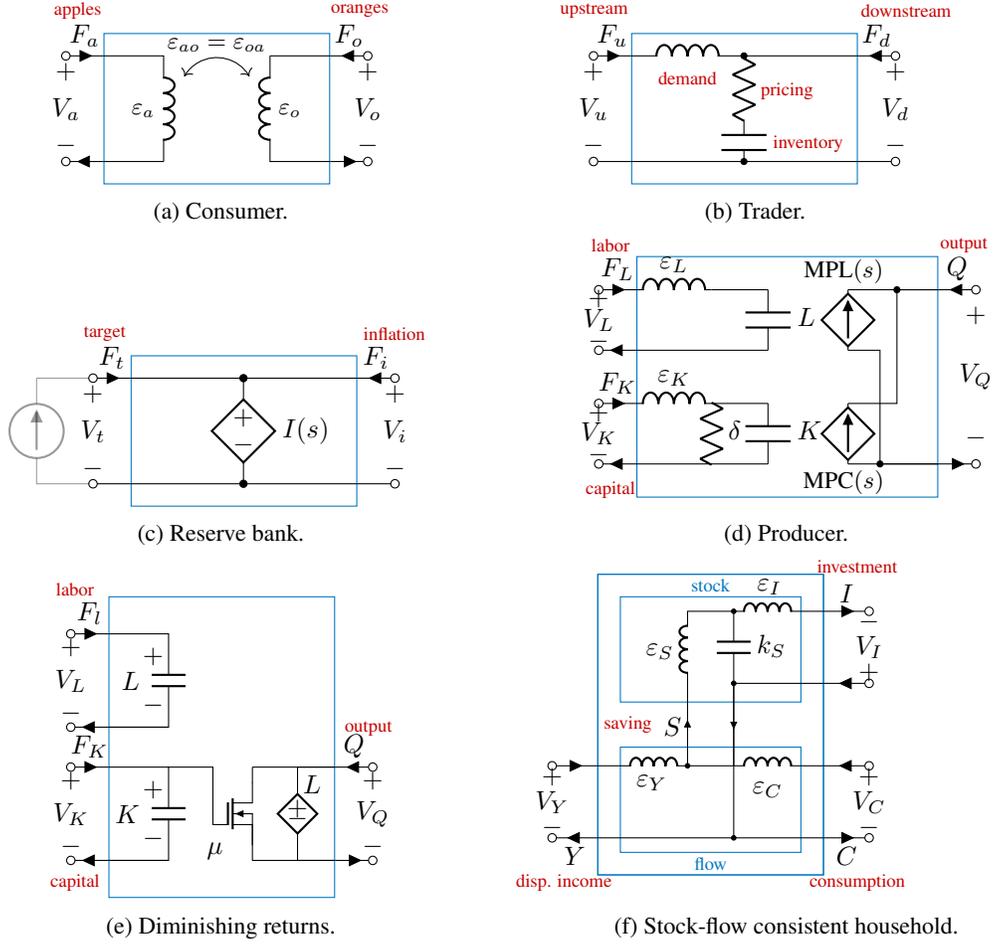
\begin{figure*}[ht]
    \centering
    \begin{subfigure}{0.4\textwidth}
        \centering
           \begin{circuitikz}[scale=1]
        \newcommand{\w}{3cm}
        \newcommand{\h}{2cm}

         \node[draw,agentcolor, ,rectangle, minimum width=\w, minimum height=\h] (A) at (0,0) {};
            \draw (A.155) to[short,-o,i<_={$F_a$}] ++(-0.5cm,0) coordinate(A1){};
            \draw (A.205) to[short,-o,i={\scriptsize \textcolor{white}{.}}] ++(-0.5cm,0) coordinate(A2){};
            \draw (A.25)  to[short,-o,i<=$F_o$] ++(0.5cm,0) coordinate(A3){};
            \draw (A.335) to[short,-o,i_={\scriptsize \textcolor{white}{.}}] ++(0.5cm,0) coordinate(A4){};
            \draw (A1) to[open,v=$V_{a}$] (A2);
            \draw (A3) to[open,v=$V_{o}$] (A4);
            \node[left of=A, xshift=-.85cm,yshift=1.3cm] {\scriptsize \textcolor{myred}{apples}};
            \node[right of=A, xshift=.9cm,yshift=1.3cm] {\scriptsize \textcolor{myred}{oranges}};

            \ctikzset{bipoles/length=1cm}
            \draw (A.205) to[] ++(.8cm,0) coordinate(C);
            \draw (A.155) to[] ++(.8cm,0) to[L,l_={\footnotesize $\varepsilon_{a}$}] (C);
            \draw (A.25)  to[] ++(-.8cm,0) coordinate(B);
            \draw (A.335) to[] ++(-.8cm,0) to[L,l_={\footnotesize $\varepsilon_{o}$}] (B);

            \draw[<->] ($(A.center) + (0cm,0.18cm)$) 
    ++(30:0.5cm) arc[start angle=30, end angle=150, radius=0.5cm] ;
    \node at ($(A.center) + (0cm,0.85cm)$) {\footnotesize $\varepsilon_{ao}=\varepsilon_{oa}$};
      \end{circuitikz}  
        \caption{Consumer.}
        \label{fig:Consumer}
    \end{subfigure}
    \begin{subfigure}{0.45\textwidth}
        \centering
    \begin{circuitikz}
       \newcommand{\h}{2cm}
        \newcommand{\w}{1.5*\h}
          \node[draw,agentcolor, rectangle, minimum width=\w, minimum height=\h] (A) at (0,0) {};

            \draw (A.155) to[short, i<_=$F_u$] ++(-\w/6,0) node[ocirc](A1){};
            \draw (A.205) to[short,] ++(-\w/6,0) node[ocirc](A2){};
            \draw (A.25)  to[short, i<^=$F_d$] ++(\w/6,0) node[ocirc](A3){};
            \draw (A.335) to[short,] ++(\w/6,0) node[ocirc](A4){};
            \draw (A1) to[open, v=$V_u$] (A2);
            \draw (A3) to[open, v=$V_d$] (A4);

        \node[left of=A, xshift=-1cm,yshift=1.3cm] {\scriptsize \textcolor{myred}{upstream}};
            \node[right of=A, xshift=1.15cm,yshift=1.3cm] {\scriptsize \textcolor{myred}{downstream}};
      \ctikzset{bipoles/length=1cm}
         \draw  (A.155)  to[L,l_={\scriptsize \textcolor{myred}{demand}}] ++(\w/2,0) coordinate(a1)  to[] (A.25)
         (A.205) to  ++(\w/2,0) coordinate(a2) to (A.335)
         (a1) to[R={\scriptsize \textcolor{myred}{pricing}},*-] ++(0,-.3*\w) to[C={\scriptsize \textcolor{myred}{inventory}},-*] (a2)
                ;

    \end{circuitikz}
    \caption{Trader.}
    \label{fig:trader}
    \end{subfigure}
    \begin{subfigure}{0.45\textwidth}
        \centering
    \begin{circuitikz}
       \newcommand{\h}{2cm}
        \newcommand{\w}{1.5*\h}
          \node[draw,agentcolor, rectangle, minimum width=\w, minimum height=\h] (A) at (0,0) {};

            \draw (A.155) to[short, i<_=$F_t$] ++(-\w/6,0) node[ocirc](A1){};
            \draw (A.205) to[short,] ++(-\w/6,0) node[ocirc](A2){};
            \draw (A.25)  to[short, i<^=$F_i$] ++(\w/6,0) node[ocirc](A3){};
            \draw (A.335) to[short,] ++(\w/6,0) node[ocirc](A4){};
            \draw (A1) to[open, v=$V_t$] (A2);
            \draw (A3) to[open, v=$V_i$] (A4);

            \node[left of=A, xshift=-.85cm,yshift=1.3cm] {\scriptsize \textcolor{myred}{target}};
            \node[right of=A, xshift=1cm,yshift=1.3cm] {\scriptsize \textcolor{myred}{inflation}};

      \ctikzset{bipoles/length=1.2cm}
         \draw  (A.155)  to[short,-*] ++(\w/2,0) coordinate(a1)  to[] (A.25)
         (A.205) to[short,-*] ++(\w/2,0) coordinate(a2) to  (A.335)
         (a1) to[cV,l=$I(s)
         $] (a2);
         \draw [opacity=.4]
         (A1) to ++(-\w/4,0) coordinate (c)
         (A2) to ++(-\w/4,0) to[american current source] (c);    
    \end{circuitikz}
        \caption{Reserve bank.}
        \label{fig:bank}
    \end{subfigure}
    \begin{subfigure}{0.45\textwidth}
        \centering

   \begin{circuitikz}
        \newcommand{\w}{4cm}
        \newcommand{\h}{3.2cm}

         \node[draw,agentcolor, rectangle, minimum width=\w, minimum height=\h] (A) at (0,0) {};
            \draw (A.150) to[short,-o,i<_=$F_L$] ++(-0.5cm,0) coordinate(A1){};
            \draw (A.170) to[short,-o,i=\textcolor{white}{.}] ++(-0.5cm,0) coordinate(A2){};
            \draw (A.190) to[short,-o,i<_=$F_K$] ++(-0.5cm,0) coordinate(A5){};
            \draw (A.210) to[short,-o,i=\textcolor{white}{,}] ++(-0.5cm,0) coordinate(A6){};
            \draw (A.30)  to[short,-o,i<=$Q$] ++(0.5cm,0) coordinate(A3){};
            \draw (A.330) to[short,-o,i_=\textcolor{white}{.}] ++(0.5cm,0) coordinate(A4){};
            \draw (A1) to[open,v=$V_L$] (A2);
            \draw (A5) to[open,v=$V_K$] (A6);
            \draw (A3) to[open,v=$V_{Q}$] (A4);

            \node[left of=A, xshift=-1.35cm,yshift=-1.5cm] {\scriptsize \textcolor{myred}{capital}};
            \node[left of=A, xshift=-1.35cm,yshift=1.75cm] {\scriptsize \textcolor{myred}{labor}};
            \node[right of=A, xshift=1.35cm,yshift=1.75cm] {\scriptsize \textcolor{myred}{output}};

            \ctikzset{bipoles/length=1cm}
            \draw (A.170) to[] ++(1cm,0) coordinate(C) to ++(.75cm,0) coordinate(P);
            \draw (A.150) to[L,l=$\varepsilon_L$] ++(1cm,0) coordinate(Q);
            \draw (Q) to++(.75cm,0) coordinate (W);
            \draw (P) to[C,l_=$L$] (W); 
            \draw (A.210) to[] ++(1cm,0) coordinate(C) to++(.75cm,0) coordinate (D);
            \draw (A.190) to[L,l=$\varepsilon_K$] ++(1cm,0) coordinate (X) to[R=$\delta$] (C);
            \draw (X)to++(.75cm,0) coordinate (Y);
            \draw (D) to[C,l_=$K$] (Y);
            \draw (A.330)  to[] ++(-1.2cm,0) coordinate(B);
            \draw (A.330)  to[short,-*] ++(-.775cm,0) coordinate(G);
            \coordinate (E) at ($(A.10) - (1.2cm,0)$);
            \coordinate (F) at ($(A.350) - (1.2cm,0)$); 
            \draw (A.30) to[] ++(-1.2cm,0) coordinate(ee);
            \draw (E) to[cI,l_=] (ee);
            \draw (G) |- (E);
            \draw (A.30) to[short,-*] ++(-.55cm,0) |- (F);
            \draw (B) to[cI] (F);

            \node at (0.75,1.4) {\small $\MPL(s)$};
            \node at (0.75,-1.4) {\small $\MPC(s)$}; 
    \end{circuitikz}
    \caption{Producer.}
    \label{fig:Multivarprod}
    \end{subfigure}
        \begin{subfigure}{.45\textwidth}
    \centering
    \begin{circuitikz}
        \newcommand{\w}{3cm}
        \newcommand{\h}{4cm}

         \node[draw,agentcolor, rectangle, minimum width=\w, minimum height=\h] (A) at (0,0) {};
            \draw (A.135) to[short,-o,i<_=$F_l$] ++(-0.5cm,0) coordinate(A1){};
            \draw (A.170) to[short,-o,i=\textcolor{white}{,}] ++(-0.5cm,0) coordinate(A2){};
            \draw (A.190) to[short,-o,i<_=$F_K$] ++(-0.5cm,0) coordinate(A3){};
            \draw (A.225) to[short,-o,i=\textcolor{white}{,}] ++(-0.5cm,0) coordinate(A4){};
            \draw (A.350)  to[short,-o,i<=$Q$] ++(0.5cm,0) coordinate(A5){};
            \draw (A.315) to[short,-o,i_={\textcolor{white}{.}}] ++(0.5cm,0) coordinate(A6){};
            \draw (A1) to[open,v=$V_L$] (A2);
            \draw (A3) to[open,v=$V_K$] (A4);
            \draw (A5) to[open,v=$V_Q$] (A6);
          

            \ctikzset{bipoles/length=.75cm}
            \draw (A.170) to[]  ++(.8cm,0) coordinate(X);
            \draw (A.135) to[] ++(.8cm,0) coordinate(Z) to[C, v_=$L$] (X) to (A.170) ;
            \draw (A.225) to[] ++(.8cm,0) coordinate(D);
            \draw (A.190) to[] ++(.8cm,0) coordinate(E) to[C, v=$K$] (D) to (A.225) ;
            \draw (A.350) to[] ++(-1.1cm,0) coordinate (B);
            \draw (A.315)  to[] ++(-1.1cm,0) to [Tnigfetd,n=mos2] (B) ;
            \draw (A.315) to[] ++(-.5cm,0) coordinate (Q);
            \draw (A.350)  to[] ++(-.5cm,0) to [cV] (Q) ;
            \draw (mos2.G) |- (E) (mos2.B) node[anchor=west]{};
            \node[below of=mos2, xshift=-.5cm, yshift=.5cm] {$\mu$};
            \node[xshift=1.2cm,yshift=-.5cm]{$L$};
            \node[left of=A, xshift=-.95cm,yshift=-1.8cm] {\scriptsize \textcolor{myred}{capital}};
            \node[left of=A, xshift=-.95cm,yshift=2.1cm] {\scriptsize \textcolor{myred}{labor}};
            \node[right of=A, xshift=.95cm,yshift=.25cm] {\scriptsize \textcolor{myred}{output}};

    \end{circuitikz}
    \caption{Diminishing returns.}
    \label{fig:dimprod}
    \end{subfigure}
  \begin{subfigure}{0.45\textwidth}
    \begin{circuitikz}
        \newcommand{\w}{3cm}
        \newcommand{\h}{4cm}
 
         \node[draw,agentcolor, rectangle, minimum width=\w, minimum height=\h] (A) at (0,0) {};
 
         \node[draw,agentcolor, rectangle, minimum width=0.8*\w, minimum height=0.35*\h] (B) at ([yshift=0.25*\h]A) {};
 
         \node[draw,agentcolor, rectangle, minimum width=0.8*\w, minimum height=0.35*\h] (C) at ([yshift=-0.25*\h]A) {};
 
         \node[draw,agentcolor, rectangle, minimum width=\w, minimum height=\h] (A) at (0,0) {};

            \draw (A.20) to[short,-o,i<_=\textcolor{white}{ }] ++(0.6cm,0) coordinate(A1){};
            \draw (A.45) to[short,-o,i=$I$] ++(0.6cm,0) coordinate(A2){};
            \draw (A.200) to[short,-o,i<_=\textcolor{white}{,}] ++(-0.6cm,0) coordinate(A3){};
            \draw (A.225) to[short,-o,i=$Y$] ++(-0.6cm,0) coordinate(A4){};
            \draw (A.340)  to[short,-o,i<=\textcolor{white}{.}] ++(0.6cm,0) coordinate(A5){};
            \draw (A.315) to[short,-o,i_=$C$] ++(0.6cm,0) coordinate(A6){};
            \draw (A1) to[open,v=$V_{I}$] (A2);
            \draw (A3) to[open,v=$V_{Y}$] (A4);
            \draw (A5) to[open,v=$V_C$] (A6);

            \ctikzset{bipoles/length=.75cm}
            \draw (A.45) to ++(-0.1*\w,0) to[L,mirror] node[above,xshift=4mm,yshift=1mm] {$\varepsilon_{I}$} ++(-0.9cm,0) coordinate(A4);
            \draw (A.200) to ++(0.1*\w,0) to[L] node[below,xshift=-4mm,yshift=0mm] {$\varepsilon_{Y}$} ++(0.9cm,0) coordinate(A2);
            \draw (A.340) to ++(-0.1*\w,0) to[L,mirror,l=$\varepsilon_{C}$] ++(-0.9cm,0) to [short,-*] (A2);
            \draw (A.315) to ++(-1.2cm,0) [short,*-] coordinate(A3) to (A.225);
            \draw (A.20) to ++(-0.9cm-0.1*\w,0) coordinate(A5) -| (A3);
            \draw (A4) to[C,mirror,l=$k_S$,*-*] (A5);
            \draw (A2) to[short,i>_=\text{ }] coordinate(A6) ++(0,1.1cm) to[L,l=$\varepsilon_{S}$] ++(0,0.9cm) |- (A4);
            \node[xshift=-0.2cm,yshift=-0.3cm] at (A6) {$S$};
            \draw (A5) to[short,i>_=\text{ }] ++(0,-1.1cm);

            \node[left of=A, xshift=-.95cm,yshift=-2.1cm] {\scriptsize \textcolor{myred}{disp. income}};
            \node[right of=A, xshift=.95cm,yshift=2.1cm] {\scriptsize \textcolor{myred}{investment}};
            \node[right of=A, xshift=0.95cm,yshift=-2.1cm] {\scriptsize \textcolor{myred}{consumption}};
            \node[left of=A, xshift=-0.1cm,yshift=-0cm] {\scriptsize \textcolor{myred}{saving}};
            \node[yshift=-1.85cm] {\scriptsize \textcolor{agentcolor}{flow}};
            \node[yshift=1.85cm] {\scriptsize \textcolor{agentcolor}{stock}};
    \end{circuitikz}
    \caption{Stock-flow consistent household.}
    \label{fig:Household}
  \end{subfigure}     
    \caption{Example  agent circuits.}
    \label{fig:exmpagents}
\end{figure*}

In this section, we illustrate how economic circuit theory determines the dynamical behavior of an agent as it appears at the agent's ports.  In the following subsections, we do this by means of the example agents in Figure \ref{fig:exmpagents} and refer the reader to \cite{Hutters2024} for further examples.   


\subsubsection*{(a) Consumer with Substitute Goods}
\label{sec:ConsumerWithSubstituteGoods}
Figure \ref{fig:Consumer} represents a consumer who demands apples and oranges at quantities $F_a$ and $F_o$, respectively.  Because the consumer is active in two markets simultaneously, its demand schedule is implemented as two inductors configured with a mutual inductance.  In this way, the $\Y$-parameter matrix in (\ref{eq: crosselasticity}) features the principal price elasticities on the diagonal and the cross-elasticities on the off-diagonal elements \cite{Hutters2024}. 
Assuming that the incentives $V_a$ and $V_o$ are exogenous, the consumer's demand is given by a $\Y$-parameter model:

\bigskip
\smallskip

\begin{equation}
\Y = \frac{1}{s}
             {\begin{pmatrix}
        \tikzmarknode{aa}{\varepsilon_{aa} } &  \tikzmarknode{ao}{\varepsilon_{ao} } \\
         \tikzmarknode{oa}{\varepsilon_{oa} } &  \tikzmarknode{oo}{\varepsilon_{oo} } \\
        \end{pmatrix}}
            \label{eq: crosselasticity}
 \end{equation}
         \begin{tikzpicture}[overlay,remember picture]
\draw[<-] (aa.north) --++ (0,0.5) node[ left,  align=right] {\footnotesize price elasticity apples};
\draw[<-] (ao.north) --++ (0,0.5) node[ right,text width=3cm] {\footnotesize cross-elasticity orange price  on apples};
\draw[<-] (oa.south) --++ (0,-0.7) node[ left, text width=3cm,  align=right] {\footnotesize cross-elasticity apple price on oranges};
\draw[<-] (oo.south) --++ (0,-0.7) node[ right,] {\footnotesize price elasticity oranges};

\end{tikzpicture}

\vspace{8mm}

The frequency-domain integrator $1/s$ transforms the exogenously imposed incentives into the agent's endogenous prices for the goods.  The elasticities then transform these into the demanded flows $F_a$ and $F_o$ that make up the consumer's response. The principal elasticities determine the influence of the good's own price on the demand, and the cross elasticities determine the influence of the other goods' price.
(See, further, \cite{Mendel2023} and \cite{Hutters2024}.)

\subsubsection*{(b) Trader as RLC circuit}
\label{sec:TraderAsRLCCircuit}
Figure \ref{fig:trader} contains a model of a trader, or other type of intermediary, for a particular good.  It appears in Figure \ref{fig:MatchingNetwork} of Section \ref{sec:bullwhip} in the context of a supply chain, where it moves goods from upstream to downstream.   The trader maintains a price in the inductor, an inventory in the capacitor, and generates trade friction in the resistor.   

The choices for the parameter models are given in Figure \ref{tab:TwoPort}.  In a supply chain, $\T$ parameters are the natural choice. In a supply-driven scenario, the upstream port is the input, and the trader responds to the exogenously imposed incentives and flows of the supplier. In a demand-driven scenario, it is the consumers who apply the incentives and flows at the downstream port, and the trader passes these on to the supplier.   

In either case, any goods that are not cleared as they move along the supply chain flow in or out of the inventory held in the capacitor.  Any mismatch between the incentives at the ports is balanced by the incentives that drive the price on the inductor (see also \cite{Hutters2024}).

\subsubsection*{(c) Reserve Bank as Controlled Source}
\label{sec:ReserveBankAsControlledSource}
In Figure \ref{fig:bank}, we use a current-controlled voltage source to model an institution, such as a Federal Reserve Bank, that is tasked with controlling the rate of interest.  The committee, such as the Federal Open Market Committee (FOMC), provides it a target rate $F_t$.  The observed actual rate $F_a$ is also imposed exogenously.  The bank responds by communicating incentives to the economy to adjust the actual rate, and to the committee to adjust the target.  This leads to a $\Z$-parameter model for the execution of monetary policy.   
\begin{equation}
\begin{pmatrix}
V_t\\
V_a
\end{pmatrix}
= I(s)
\begin{pmatrix}
1 & -1\\
1 & -1
\end{pmatrix}
\begin{pmatrix}
F_t\\
F_a
\end{pmatrix}.
\end{equation}
The matrix is chosen so that the bank reacts to the spread,  i.e., the deviation $F_t-F_a$ of the actual rate from the target.  We refer to $I(s)$ as the policy function.  We let it be a function of the complex frequency $s$.  This allows the bank to let its policy depend on the discount rate and the frequency of the seasonality.  Whenever convenient, this can be transformed to the time domain to describe the dynamical behavior.  

With this model, the bank can be thought of as a feedback controller.  The target rate serves as the setpoint,  the spread serves as the error, and the policy function serves as the control law. An interesting example is to set the policy function equal to the control law of a PID controller, i.e., 
\begin{equation}
  I(s) = K_p+\frac{K_i}{s}+K_d s 
\end{equation}
The proportional gain $K_p$ measures how aggressively the bank should act at the moment.   The integral gain $K_i$ does this for the  drift that accumulates over time.  The derivative gain $K_d$ allows the bank to anticipate and can be useful when there is a tendency to strongly overshoot or oscillate.   

\subsubsection*{(d) Marginal Product with Controlled Current Sources}
\label{sec:MarginalProduct}
Economists use production functions to determine the marginal product of the factors of production.  Typically, a function of the form $Q = Q( K, L)$ is chosen.  The quantity of output $Q$ is a flow of goods and the inputs $K$ and $L$ are stocks representing the physical capital and labor, respectively (see, e.g., \cite{Varian2010}).  The marginal products of capital and labor are then defined, respectively,  as follows:
\begin{equation}
 \MPC = \pdv{Q}{K}, \qquad  \MPL = \pdv{Q}{L} .
 \label{eq:MarginalProduct}
\end{equation}

Figure  \ref{fig:Multivarprod} contains the design of a multiport agent representing a producer or productivity sector of the economy.   It has three ports: two for the factor inputs and one for the output.  The inductors on the inputs represent the demand for the factor flows, i.e., hiring of workers and acquisition rates of capital equipment. The capacitors record the total stock of the factors, and a parallel resistor on the capital stock represents its depreciation.  
The production function is encoded in the two controlled current sources. When connected in parallel, the circuit gives the following function:
\begin{equation}
    Q = \MPC(s) K + \MPL(s) L .
\end{equation}

From an economic perspective, this is a linear production function with coefficients equal to the marginal productivities.  From a network perspective, the productivities are the elements of the agent's parameter matrix.   They are  the transfer functions between the factor inputs and the output of goods:
\begin{equation}
    \MPC(s) = \frac{Q(s)}{K(s)}  \qquad  \MPL(s) = \frac{Q(s)}{L(s)} .
\label{eq:TransferFunctionMP}
\end{equation}
Unlike the productivities in (\ref{eq:MarginalProduct}), the transfer functions in (\ref{eq:TransferFunctionMP}) depend on the complex frequency $s$.  For economic applications, their dependence on the frequency of the seasonality and the own rate of discount provides an important generalization in our opinion.   The poles of the transfer function, in particular, are critical to analyzing the dynamic response of the producer when the inputs are not static flows. 
In Section \ref{sec:bullwhip}, we show the use of Bode plots to compare the productivity as it depends on the frequency of the seasonality of the demand.   

\subsubsection*{(e) Diminishing Returns with Field-Effect Transistors}

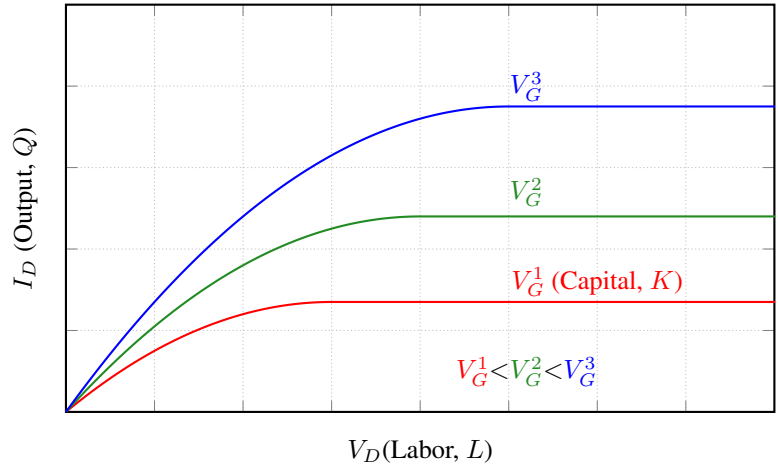
\begin{figure}[ht]
        \centering
      \begin{tikzpicture}

  \def\Vth{1.0}   
  \def\k{0.6}     

  \begin{axis}[
    width=11cm,
    height=7cm,
    xlabel={$V_{D}$(Labor, $L$)  },
    ylabel={$I_D$ (Output, $Q$)},
    xmin=0, xmax=4,
    ymin=0, ymax=5,
    grid=both,
    major grid style={densely dotted},
    minor grid style={densely dotted},
    thick,
    domain=0:5,
    samples=200,
    xticklabels={},
    yticklabels={}
  ]

    \newcommand{\IdCurve}[2]{%
      \addplot[color=#2]
      { (#1 <= \Vth)
          ? 0
          : ( x <= (#1-\Vth)
              ? \k*(2*(#1-\Vth)*x - x^2)
              : \k*(#1-\Vth)^2
            )
      };
    }

    \IdCurve{2.5}{red}
    \IdCurve{3}{ForestGreen}
    \IdCurve{3.5}{blue}

    \node[red]   at (axis cs:3,1.6)  {$V_{G}^1$ (Capital, $K$)};
    \node[ForestGreen] at (axis cs:2.6,2.7)  {$V_{G}^2$};
    \node[blue] at (axis cs:2.6,4.0) {$V_{G}^3$};
    \node at (axis cs:2.6,.5) {\textcolor{red}{$V_{G}^1$}$<$\textcolor{ForestGreen}{$V_{G}^2$}$<$\textcolor{blue}{$V_{G}^3$}};
  \end{axis}
\end{tikzpicture}
    \caption{MOSFET output characteristics \cite{Carson1922} as production functions with diminishing returns \cite{zivot_production_function_2025}.}
    \label{fig:ProductionTransistorCharacteristics}
\end{figure}

Typically, production functions are assumed to be nonlinear due to the diminishing marginal productivity of inputs (see e.g. \cite{Varian2010}).
As we show in Figure \ref{fig:ProductionTransistorCharacteristics}, the characteristics of such production functions appear similar to the drain current of a MOSFET when graphed.

Figure \ref{fig:dimprod} contains an example circuit that uses a MOSFET to implement a rudimentary production function with diminishing productivity of labor and a minimum required capital.  
Like the network in Figure \ref{fig:Multivarprod}, the multiport concerns labor, capital, and output.

The MOSFET is configured so that its drain current represents the output $Q$. The capacitance of the capital stock is set to unity so that the gate voltage of the MOSFET equals the capital stock itself, i.e., $V_K = K$.  Similarly, a controlled source then sets its drain voltage equal to the labor $L$.  This results in the following nonlinear production function:
\begin{equation}
    Q = \mu\left((
    K-K_{th})L-\frac{L^2}{2}\right),
\end{equation} 
which corresponds to the DC drain current of the MOSFET.
The transconductance $\mu$ of the MOSFET serves to quantify the factor productivity.  It is straightforward to check that the function is diminishing in labor.  For capital, it requires a minimum amount $K_{th}$, given by the threshold gate-source voltage, for production to start.

\subsubsection*{(f) Stock-Flow Consistent Household}
Stock-flow consistent modeling is a relatively new technique in economics that enforces a strict accounting for all relevant flows and their stocks \cite{Caiani2016}.  Using multiport networks, stock-flow consistency is automatically enforced. In \cite{Hutters2024}, we show how Kirchhoff's current law enforces it at the economic circuit level.  For multiport networks, it is the port condition, as described in Section \ref{sec:InteractionsWithPorts}, that guarantees stock-flow consistency among the multiport agents in a network.

The example in Figure \ref{fig:Household} models a household as a combination of a balance sheet 2-port and a flow statement 3-port.  It is taken from \cite{Berend} where it forms part of a stock-flow consistent multiport model of an economy.  The flow statement has a port for disposable income $Y$, consumption $C$, and saving $S$.  These flows are consistent with the GDP identity $Y=C+S$ that forms the household's budget constraint.  Saving $S$ enters the balance sheet, which also has a port for investment $I$.  The difference $S-I$, which the household holds as cash on hand, is kept in the capacitor. By combining a flow statement with a balance sheet, the household multiport is rendered stock-flow consistent over both income and capital flows.   

The inductors add dynamic behavior to the household.  At the external facing ports, they communicate the household's desire for income and consumption, as well as the incentive required to invest. At the internal saving port, the household consolidates its books, and the inductor determines the transactional motive for holding cash.   

\subsection{Agents as equivalent 2-ports}
\label{sec:AgentsAsEquivalent2Ports}
It is often convenient to add a level of abstraction between the multiport agent and its internal circuit.   For this, we exploit the concept of an equivalent circuit.  This is a circuit that contains a component for each parameter in the parameter matrix.  It allows a designer to rapidly establish a working network that exhibits the correct dynamical behavior, postponing the precise implementation of the circuit to a later stage.

In Table \ref{tab:equivagents}, we present the equivalent circuits for the 2-port parameter models introduced in Table \ref{tab:TwoPort}.  These contain four components.  Two are passive (rectangle shapes), i.e., they do not require surplus to operate. They correspond to the diagonal entries of the matrix, where the principal effects of the operator are determined.  The other two are active (diamond shapes), i.e., they do require surplus, and correspond to the antidiagonal elements which determine the cross effects.

Their placement in the matrix suggests certain economic interpretations for the parameters.  These are indicated in the last column of Table \ref{tab:equivalentmodels}.  In addition to the familiar price elasticity and its inverse, the price inelasticity, we recognize two other concepts.    
Price transmission is a concept in the economic literature (see, e.g., \cite{von2021price}), where the effect of one price on the other is quantified.  This is, for instance, the case for $\T_{11}$ parameter of the transmission model in Table  \ref{tab:equivagents}.  It is a dimensionless quantity.  A related concept, similarly dimensionless, that emerges in this context is what we refer to as flow throughput. In the table, the $\T_{22}$ parameter forms an example.   


\begin{table*}[]
    \caption{Parameter-equivalent agent abstractions.}
    \label{tab:equivagents}
    \centering
    \begin{tabular}{|c|c|>{\centering\arraybackslash}p{7.2cm}|}
        \hline
        \textbf{Type} & \textbf{Circuit} & \textbf{Parameter Model} \\
        \hline
        {\footnotesize Admittance}
         & \parbox[c][3cm][c]{4.5cm}{
         \includegraphics[width=\linewidth]{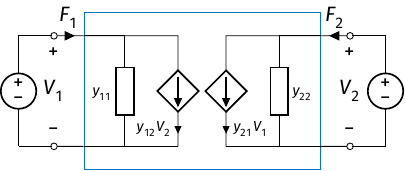}
         }
         &{\footnotesize
         $        \underbrace{\begin{pmatrix}
            \tikzmarknode{11}{\Y_{11}} & \tikzmarknode{12}{\Y_{12}}\\
           \tikzmarknode{21}{\Y_{21}} & \tikzmarknode{22}{\Y_{22}}
        \end{pmatrix}}_{\mathbf{\Y}}
         $}
         \begin{tikzpicture}[overlay,remember picture]
\draw[<-] (11.north) --++ (0,0.5) node[ left, text width=2.5cm, align=right] {\scriptsize price elasticity good 1};
\draw[<-] (12.north) --++ (0,0.5) node[ right, text width=2.5cm] {\scriptsize cross-elasticity of price 2 on good 1};
\draw[<-] (21.south) --++ (0,-0.7) node[ left, text width=2.5cm, align=right] {\scriptsize cross-elasticity of price 1 on good 2};
\draw[<-] (22.south) --++ (0,-0.7) node[ right, text width=2.5cm] {\scriptsize price elasticity good 1};

\end{tikzpicture}
        \\
        \hline
      \footnotesize  Impedance &
        \parbox[c][3cm][c]{4.5cm}{
        \includegraphics[width=\linewidth]{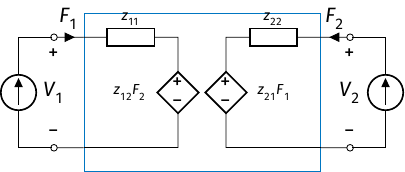}
         }
         &{\footnotesize
                  $        \underbrace{\begin{pmatrix}
            \tikzmarknode{11}{\Z_{11}} & \tikzmarknode{12}{\Z_{12}}\\
           \tikzmarknode{21}{\Z_{21}} & \tikzmarknode{22}{\Z_{22}}
        \end{pmatrix}}_{\mathbf{\Z}}
         $}
         \begin{tikzpicture}[overlay,remember picture]
\draw[<-] (11.north) --++ (0,0.5) node[ left, text width=2.5cm, align=right] {\scriptsize price elasticity good 1};
\draw[<-] (12.north) --++ (0,0.5) node[ right, text width=2.5cm] {\scriptsize cross-elasticity of price 2 on good 1};
\draw[<-] (21.south) --++ (0,-0.7) node[ left, text width=2.5cm, align=right] {\scriptsize cross-elasticity of price 1 on good 2};
\draw[<-] (22.south) --++ (0,-0.7) node[ right, text width=2.5cm] {\scriptsize price elasticity good 1};
\end{tikzpicture}
        \\
        \hline
       \footnotesize Transmission&
         \parbox[c][3cm][c]{4.5cm}{   
        \includegraphics[width=\linewidth]{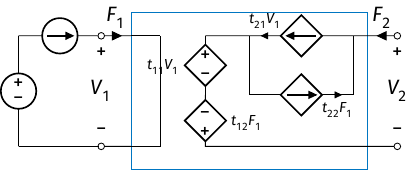}
         }
        &{\footnotesize
                 $        \underbrace{\begin{pmatrix}
            \tikzmarknode{11}{\T_{11}} & \tikzmarknode{12}{\T_{12}}\\
            \tikzmarknode{21}{\T_{21}} &  \tikzmarknode{22}{\T_{22}}
        \end{pmatrix}}_{\mathbf{\T}} 
         $}
         \begin{tikzpicture}[overlay,remember picture]
\draw[<-] (11.north) --++ (0,0.5) node[ left, text width=2.5cm, align=right] {\scriptsize price transmission};
\draw[<-] (12.north) --++ (0,0.5) node[ right, text width=2.5cm] {\scriptsize cross-inelasticity of price 1 on good 2};
\draw[<-] (21.south) --++ (0,-0.7) node[left, text width=2.5cm, align=right] {\scriptsize cross-elasticity of price 1 on good 2};
\draw[<-] (22.south) --++ (0,-0.7) node[right , text width=2.5cm] {\scriptsize flow throughput};

\end{tikzpicture}
        
         \\
         \hline
        \small Hybrid  
         &\parbox[c][3cm][c]{4.5cm}{
          \includegraphics[width=\linewidth]{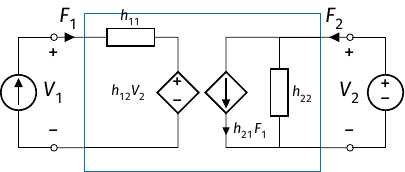}
         }
         &{\footnotesize
                          $        \underbrace{\begin{pmatrix}
            \tikzmarknode{11}{\HH_{11}} & \tikzmarknode{12}{\HH_{12}}\\
            \tikzmarknode{21}{\HH_{21}} &  \tikzmarknode{22}{\HH_{22}}
        \end{pmatrix}}_{\mathbf{\HH}}
         $}
         \begin{tikzpicture}[overlay,remember picture]
\draw[<-] (11.north) --++ (0,0.5) node[ left, text width=2.5cm, align=right] {\scriptsize price elasticity good 1};
\draw[<-] (12.north) --++ (0,0.5) node[ right, text width=2.5cm] {\scriptsize price transmission};
\draw[<-] (21.south) --++ (0,-0.7) node[left, text width=2.5cm, align=right] {\scriptsize flow throughput};
\draw[<-] (22.south) --++ (0,-0.7) node[ right, text width=2cm] {\scriptsize price elasticity good 2};
\end{tikzpicture}
         
        \\
        \hline
    \end{tabular}
\end{table*}

\section{Hierarchical Modeling and Aggregation}
\label{sec:InteractionsInterconnections}
In the introduction, we already considered the problems in determining the emergent dynamics for aggregate agents, as they are documented in both the macroeconomic and agent-based economic traditions.  In this section, we show how the multiport framework provides a structured and mathematically rigorous approach to these problems.  

In multiport theory, the representation of a model at a higher level of abstraction is created by interconnecting several multiports into a single aggregate multiport. As long as the port condition is fulfilled,  this process leads to a valid multiport network for the aggregate agent \cite{alexander2012fundamentals}.  Therefore, its dynamical behavior remains emergent from underlying circuit-level descriptions. 

The aggregation process can be repeated to create a hierarchy of levels of abstraction.   The port condition guarantees that the internal consistency of the model is preserved and the correct dynamical behavior emerges, independent of the heterogeneity of the constituent agents.

\begin{table}[h]
    \centering
     \caption{Equivalent operators for various modes of interaction and their interconnection topologies  \cite{alexander2012fundamentals}.}
    \label{tab:equivalentmodels}
    \renewcommand{\arraystretch}{1.2} 
    \begin{tabular}{llcc}
        \toprule
        \multicolumn{2}{c}{\textbf{Topology}} &  \multicolumn{2}{c}{\textbf{Algebra}}  \\
        \textbf{Interaction} & \textbf{Interconnection} & \textbf{Aggregate} & \textbf{Representative}\\ \midrule
        Competition & Parallel & $
        \sum_{k=1}^n \Y_k$  & $\frac{1}{n} \sum_{k=1}^n \Y_k$  \\[5pt]
        Cooperation & Series & $\sum_{k=1}^n \Z_k$  & $\frac{1}{n}  \sum_{k=1}^n \Z_k$ \\ [5pt]
        Franchise & Series-Parallel & $\sum_{k=1}^n \HH_k$ & $\frac{1}{n}\sum_{k=1}^n \HH_k$
        \\ [5pt]
                Chain & Cascade & $\prod_{k=1}^n \T_k
        $ & $\left( {\prod_{k=1}^n \T_k} \right)^{\frac{1}{n}}
        $ \\[5pt]
        Cartel & Parallel-Series & $\sum_{k=1}^n \G_k$ & $ \frac{1}{n}\sum_{k=1}^n \G_k$
           \\ \bottomrule
    \end{tabular}
\end{table}

\subsection{The Topology and Parameters of Interaction Modes}
\label{sec:TopologyOfInteractions}

Multiport networks allow for a topological visualization of the economic interactions among agents. The particular mode in which agents interact determines how they are aggregated into a single aggregate agent. Each interaction mode corresponds to a particular topology in which the multiports are interconnected to create a single multiport.   

For 2-ports, we distinguish between the five modes of interaction listed in the first column of Table \ref{tab:equivalentmodels}.  The corresponding interconnection topologies are listed in the second column of the table.  

In the third and fourth columns, we list the formulae for the parameter models of the corresponding aggregate agent and representative agent, respectively. They are particularly intuitive.   Aggregation is done using matrix addition, except for the chaining parameters, which are multiplied. The representative agent is found by averaging: a geometric average for chains and a straightforward arithmetic average for all others.   These models provide the emergent dynamical behavior at the aggregate level of abstraction in terms of the behavior at the lower level of abstraction.  

\begin{figure*}[ht!]
    \centering
        \begin{subfigure}{.48\textwidth}
        \centering
\includegraphics[width=\textwidth]{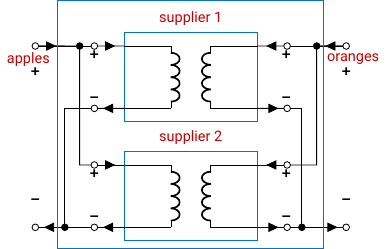}
    \caption{Competition between two suppliers operating in both the apple and orange markets (parallel).}
    \label{fig:parallel}
    \end{subfigure}\quad
    \begin{subfigure}{.48\textwidth}
        \centering        
    \includegraphics[width=\textwidth]{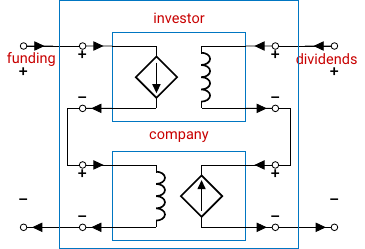}
    \caption{Cooperation between an investor providing funding and a company returning dividends (series).}
    \label{fig:series}
    \end{subfigure}
    \begin{subfigure}{.48\textwidth}
            \centering
    \includegraphics[width=\textwidth]{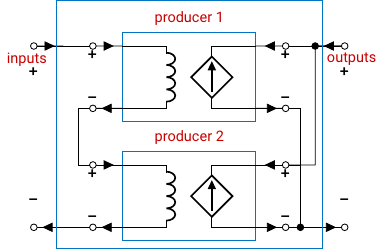}
    \caption{Franchise of  producers (series-parallel).}
    \label{fig:series-parallel}
    \end{subfigure}
    \begin{subfigure}{.48\textwidth}
            \centering
    \includegraphics[width=\textwidth]{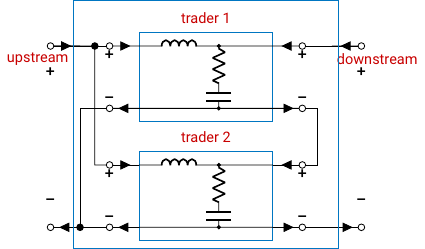}
    \caption{Cartel of traders (parallel-series).}
    \label{fig:parallel-series}
    \end{subfigure}
    \begin{subfigure}{\textwidth}
            \centering
             \begin{circuitikz}
        \newcommand{\h}{1.5cm}
        \newcommand{\w}{1.5*\h}
        \node[draw,agentcolor, rectangle, minimum width=\w, minimum height=\h] (A) at (0,0) {};

            \draw (A.155) to[short,i<_=\text{}] ++(-.5cm,0) node[ocirc](A1){};
            \draw (A.205) to[short,i=\text{}] ++(-0.5cm,0) node[ocirc](A2){};
            \draw (A.25)  to[short,i<=\text{}] ++(.5cm,0) node[ocirc](A3){};
            \draw (A.335) to[short,i_=\text{}] ++(0.5cm,0) node[ocirc](A4){};
            \draw (A1) to[open,v=\text{}] (A2);
            \draw (A3) to[open,v=\text{}] (A4);
        \node[draw, agentcolor, rectangle, minimum width=\w, minimum height=\h, right=1.5cm of A] (B) {};
            \draw (B.155) to[short,i<_=\text{}] ++(-.5cm,0) node[ocirc](B1){};
            \draw (B.205) to[short,i=\text{}] ++(-0.5cm,0) node[ocirc](B2){};
            \draw (B.25)  to[short,i<=\text{}] ++(.5cm,0) node[ocirc](B3){};
            \draw (B.335) to[short,i_=\text{}] ++(0.5cm,0) node[ocirc](B4){};
            \draw (B1) to[open,v=\text{}] (B2);
            \draw (B3) to[open,v=\text{}] (B4);     
            \draw (A3) -- (B1);
            \draw (A4) -- (B2);

        \draw (A1) to++(-.75cm,0) node[ocirc](1){};
        \draw (A2) to ++(-.75cm,0) node[ocirc](2){};
        \draw (B3) to ++(.75cm,0) node[ocirc](3){};
        \draw (B4) to ++(.75cm,0) node[ocirc](4){};

        \draw (1) to[short,i=\text{}] ++(.5cm,0);
        \draw (2) to[short,i<=\text{}] ++(.5cm,0);
        \draw (3) to[short,i=\text{}] ++(-.5cm,0);
        \draw (4) to[short,i<=\text{}] ++(-.5cm,0);
        
        \ctikzset{bipoles/length=1cm}
                \draw (A.205) to[] ++(.75cm,0) coordinate(c);
                \draw (A.155) to[] ++(.75cm,0) to[L] (c);
                \draw (A.25)  to[] ++(-.75cm,0) coordinate(b);
                \draw (A.335) to[] ++(-.75cm,0) to[cI] (b);
         \ctikzset{bipoles/length=.7cm}
         \draw  (B.155)  to[L] ++(.6*\w,0) coordinate(a1)  to[] (B.25)
         (B.205) to  ++(.6*\w,0) coordinate(a2) to (B.335)
         (a1) to[R,*-] ++(0,-.3*\w) to[C,-*] (a2)
                ;        
        \node[left of=A,yshift=.3cm,xshift=-1.85cm, myred] {\scriptsize upstream};
        \node[right of=B,yshift=.3cm,xshift=1.9cm, myred] {\scriptsize downstream};
        \node[above of=A,myred]{\scriptsize producer};
        \node[above of=B,myred]{\scriptsize trader};  

        \node[draw,agentcolor, rectangle, minimum width=3.35*\w, minimum height=1.5*\h] (A) at (.83*\w,0cm) {};
            \end{circuitikz}
            \caption{Chain with a producer and trader (cascade).}
            \label{fig:cascade}
    \end{subfigure}
    \caption{Interconnection topologies for common interaction modes. }
    \label{fig:Interconnections}
\end{figure*}
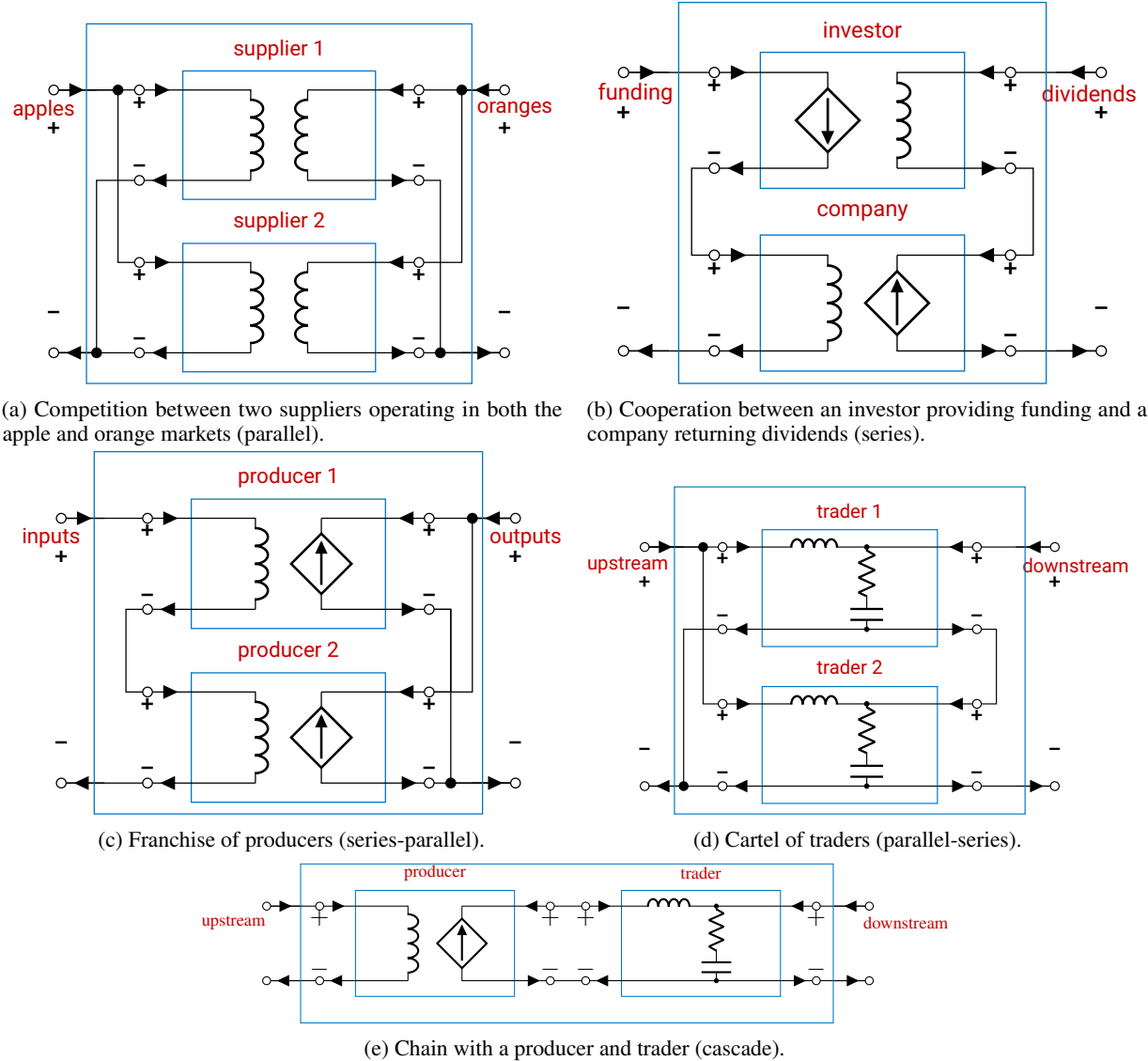

In  Figure \ref{fig:Interconnections}, we diagram the interconnection topologies for the interaction modes for pairs of 2-port agents.  In the following five subsections, respectively, we analyze the given five interaction modes and their emergent dynamics in detail.  


\subsubsection*{(a) Competition}
When competing for a limited provision $f$ of goods, the participating agents each receive a fraction of the total. In addition, they must share the available economic surplus. This means that the power $vf$, i.e., the surplus allocation rate, must also be conserved.  These two conditions imply that competing agents exhibit the same incentive.  Therefore: 
\begin{equation}
\label{eq:Competition}
f = \sum_{i=1}^m f_i \quad\quad \text{and} \quad\quad v = v_1 = \cdots = v_n .     
\end{equation}
These are the conditions for market clearing (see  \cite{Hutters2024}).

Figure \ref{fig:parallel} shows how two agents can be interconnected to enforce these equalities.  In multiport theory, this is known as a parallel interconnection.  The figure considers two suppliers competing for trade in the apple and orange markets. For each good, the market-clearing equalities have to hold separately.   

The figure shows that the aggregate supplier is also a 2-port agent, who is similarly active in both markets.  Its  $\Y-$parameter matrix is found by adding the $\Y-$parameter matrices of the individual suppliers.  A representative agent corresponds to the average of the two.

\subsubsection*{(b) Cooperation}
We consider agents to be cooperating if the goods have to go through both at the same time.  They will agree to do this when incentivized adequately, i.e., to the degree that the market clears.   This implies that
\begin{equation}
\label{eq:Cooperation}
  f = f_1 = \cdots = f_n  \quad\quad \text{and} \quad\quad  v = \sum_{i=1}^m v_i . 
\end{equation}
These equations are conjugate to those of competitive agents in (\ref{eq:Competition}).  Here, the total provided incentive, rather than the flow, is balanced among the agents, and it is the flows that that are equal. 

In Figure \ref{fig:series}, we show the interconnection that is consistent with (\ref{eq:Cooperation}).  The example concerns an investor and a stock company in a mutually beneficial relationship, with the investor supplying the funds for which the company returns a dividend stream to the investor.   

Together, they form a single aggregate agent representing the financial sector, with parameters $\Z_\text{inv} + \Z_\text{co}$.  The representative agent is found by taking the average instead.

\subsubsection*{(c) Franchise and Cartel Formation}
Agents may cooperate in some markets and compete in others.  In such a case, some of the ports are interconnected in parallel, while the others are interconnected in series.   

Figures  \ref{fig:series-parallel} and \ref{fig:parallel-series} contain two examples of common business models.  The former shows what is known as a franchise, where producers cooperate in procuring input goods but compete in supplying their outputs to the market. The latter shows a cartel formed by two traders, where they compete for the upstream supply, but they cooperate to manage the downstream distribution.   

Given the appropriate choice of parameters (see Table \ref{tab:TwoPort}), the equivalent and representative operators are determined by addition and averaging, respectively.

\subsubsection*{(d) Chaining}
In Porter's \cite{porter1985competitive} concept of a value chain, an organization is viewed as a chain of 
business units, where the output of a unit is connected to the input of the subsequent unit.  In a multiport network rendition of this, we implement the business units are 2-ports and configure them in cascade. 

This is implemented in Figure \ref{fig:cascade} for a portion of a supply chain.  A supplier sources raw material upstream, transforms these inputs into goods with value added, which are output to a trader who subsequently adds more value and outputs them downstream. 

The transformation process of a business unit is modeled by its behavioral operator, making this process fully dynamic.   The $\T$-parameters specify this in the frequency domain. (In the case study in Section \ref{sec:bullwhip}, we show how natural the latter is to analyze the bullwhip effect.)  They quantify the process of value adding by determining the relative increase in desirability $V_\text{out} / V_\text{in}$ between the input and output of each unit. 

To find the $\T$-parameters for the entire chain, we therefore multiply the $\T$-parameters of the individual units.    The representative agent is found by taking a geometric average of the latter.  
In the time domain, these are determined using the convolution operator.  This contrasts with the other interaction modes, since addition and averaging are preserved when transforming between domains.  

Rather than aggregating, it is perhaps more appropriate to refer to the process as compounding in the case of chains. Therefore, the determination of the emergent dynamics of value chains appears to be relatively more challenging.  


The rate at which the surplus is allocated over a chain can also be found using $\T$-parameters.  This works analogously to how amplifiers are chained to increase the power gain from input to output. This is useful to determine profits at each business and enhance the chain's efficiency to control overall profits.

\subsection{Aggregation and the Operator Algebra}
\label{sec:AggregationAndTheOperatorAlgebra}
The general process of determining the behavioral operator of aggregate or representative agents is done using an efficient operator algebra.   Table \ref{tab:equivalentmodels} shows it explicitly for the parameter models in the frequency domain, where the operations reduce to regular addition and multiplication.   The algebra provides designers with a structured technique for bottom-up modeling of networks.   

The procedure is to first aggregate into sub-aggregations of agents that follow the same parameter model.   Then these are aggregated again until the final network is reached.    

\begin{figure*}
    \centering
    \resizebox{\textwidth}{!}{%
    \begin{circuitikz}
        \newcommand{\h}{1.2cm}
        \newcommand{\w}{1.5*\h}
        \node[draw,PineGreen, rectangle, minimum width=\w, minimum height=\h] (A) at (0,0) {$\T_\text{OEM}$};
\node[above=1pt of A, PineGreen] {OEM};
            \draw (A.155) to[short] ++(-2*\w/8,0) node[ocirc](A1){};
            \draw (A.205) to[short,] ++(-2*\w/8,0) node[ocirc](A2){};
            \draw (A.25)  to[short,] ++(\w/8,0) node[ocirc](A3){};
            \draw (A.335) to[short,] ++(\w/8,0) node[ocirc](A4){};
            \draw (A1) to[open] (A2);
      
        \node[draw, PineGreen, rectangle, minimum width=\w, minimum height=\h, right=7/8*\w of A, yshift=\h] (B) {$\Y_B$};
            \draw (B.155) to[short] ++(-\w/8,0) node[ocirc](B1){};
            \draw (B.205) to[short] ++(-\w/8,0) node[ocirc](B2){};
            \draw (B.25)  to[short] ++(\w/8,0) node[ocirc](B3){};
            \draw (B.335) to[short] ++(\w/8,0) node[ocirc](B4){};
            \draw (B1) to[open] (B2);
            \draw (B3) to[open] (B4);     
        \node[draw, PineGreen, rectangle, minimum width=\w, minimum height=\h, below=1*\h of B] (C) {$\Y_C$};
            \draw (C.155) to[short] ++(-\w/8,0) node[ocirc](C1){};
            \draw (C.205) to[short] ++(-\w/8,0) node[ocirc](C2){};
            \draw (C.25)  to[short] ++(\w/8,0) node[ocirc](C3){};
            \draw (C.335) to[short] ++(\w/8,0) node[ocirc](C4){};
            \draw (C1) to[open] (C2);
            \draw (C3) to[open] (C4);     

            \draw (B1) to ++(-3/8*\w,0) node[ocirc](BC1){}
                       to[short,-*] ++(1/8*\w,0) |- (C1)
                  (C2) to ++(-3/8*\w,0) node[ocirc](BC2){}
                        to[short,-*] ++(2/8*\w,0) |- (B2)
                  (BC1) to ++(-1/8*\w,0) |- (A3)
                  (BC2) to ++(-1/8*\w,0) |- (A4)
                  (B3) to ++(2/8*\w,0) node[](BC3){}
                       to[short,-*] ++(-1/8*\w,0) |- (C3)
                  (C4) to[short,-*] ++(2/8*\w,0) node[](BC4){} |- (B4)

                        ;

        \node[draw, PineGreen, rectangle, minimum width=\w, minimum height=\h, right=9/8*\w of B, yshift=\h] (D) {$\Y_D$};
            \draw (D.155) to[short] ++(-\w/8,0) node[ocirc](D1){};
            \draw (D.205) to[short] ++(-\w/8,0) node[ocirc](D2){};
            \draw (D.25)  to[short] ++(\w/8,0) node[ocirc](D3){};
            \draw (D.335) to[short] ++(\w/8,0) node[ocirc](D4){};
            \draw (D1) to[open] (D2);
            \draw (D3) to[open] (D4);   
        \node[draw, PineGreen, rectangle, minimum width=\w, minimum height=\h, below=1*\h of D] (E) {$\Y_E$};
            \draw (E.155) to[short] ++(-\w/8,0) node[ocirc](E1){};
            \draw (E.205) to[short] ++(-\w/8,0) node[ocirc](E2){};
            \draw (E.25)  to[short] ++(\w/8,0) node[ocirc](E3){};
            \draw (E.335) to[short] ++(\w/8,0) node[ocirc](E4){};
            \draw (E1) to[open] (E2);
            \draw (E3) to[open] (E4);   
        \node[draw, PineGreen, rectangle, minimum width=\w, minimum height=\h, below=1*\h of E] (F) {$\Y_F$};
            \draw (F.155) to[short] ++(-\w/8,0) node[ocirc](F1){};
            \draw (F.205) to[short] ++(-\w/8,0) node[ocirc](F2){};
            \draw (F.25)  to[short] ++(\w/8,0) node[ocirc](F3){};
            \draw (F.335) to[short] ++(\w/8,0) node[ocirc](F4){};
            \draw (F1) to[open] (F2);
            \draw (F3) to[open] (F4);   

            \draw (D1) to ++(-3/8*\w,0) node[ocirc]{} to++(-1/8*\w,0) |- (B3)
                  (D1) to[short,-*] ++(-2/8*\w,0) |- (E1)
                  (F2) to ++(-3/8*\w,0)node[ocirc] {} to++(-1/8*\w,0) |- (C4)
                  (F2) to[short,-*] ++(-1/8*\w,0) |- (E2)
                  (E1) to[short,-*] ++(-2/8*\w,0) |- (F1)
                  (E2) to[short,-*] ++(-1/8*\w,0) |- (D2)
                  (D3) to[short,-*] ++(1/8*\w,0) |- (E3)
                  (E3) to[short,-*] ++(1/8*\w,0) |- (F3)
                  (F4) to[short,-*] ++(2/8*\w,0) |- (E4)
                  (E4) to[short,-*] ++(2/8*\w,0) |- (D4)
                  (D3) to[short] ++(3/8*\w,0) node[ocirc](DEF3) {}
                  (F4) to[short] ++(3/8*\w,0) node[ocirc](DEF4) {}
                        ;
        \node[draw, PineGreen, rectangle, minimum width=\w, minimum height=\h, right=9/8*\w of E, yshift=\h] (G) {$\Y_G$};
            \draw (G.155) to[short] ++(-\w/8,0) node[ocirc](G1){};
            \draw (G.205) to[short] ++(-\w/8,0) node[ocirc](G2){};
            \draw (G.25)  to[short] ++(\w/8,0) node[ocirc](G3){};
            \draw (G.335) to[short] ++(\w/8,0) node[ocirc](G4){};
            \draw (G1) to[open] (G2);
            \draw (G3) to[open] (G4);
        \node[draw, PineGreen, rectangle, minimum width=\w, minimum height=\h, below=1*\h of G] (H) {$\Y_H$};
            \draw (H.155) to[short] ++(-\w/8,0) node[ocirc](H1){};
            \draw (H.205) to[short] ++(-\w/8,0) node[ocirc](H2){};
            \draw (H.25)  to[short] ++(\w/8,0) node[ocirc](H3){};
            \draw (H.335) to[short] ++(\w/8,0) node[ocirc](H4){};
            \draw (H1) to[open] (H2);
            \draw (H3) to[open] (H4); 

        \draw (G1) to[short,-*] ++(-2/8*\w,0) |- (H1)
              (H2) to[short,-*] ++(-1/8*\w,0) |- (G2)
              (G3) to[short,-*] ++(1/8*\w,0) |- (H3)
              (H4) to[short,-*] ++(2/8*\w,0) |- (G4)
              (DEF3) to++(1/8*\w,0) |-(G1)
              (DEF4) to++(1/8*\w,0) |-(H2)
        ;
        \node[draw, agentcolor, rectangle, minimum width=\w, minimum height=\h, right=7/8*\w of H, yshift=\h] (I) {$\T_\text{EU}$};
             \node[agentcolor, above=1pt of I] {End User};   
            \draw (I.155) to[short] ++(-\w/8,0) node[ocirc](I1){};
            \draw (I.205) to[short] ++(-\w/8,0) node[ocirc](I2){};
            \draw (I1) to[open] (I2);

        \draw (G3) to ++(3/8*\w,0) node[ocirc] (GH3){} to++(1/8*\w,0)|- (I1)
              (H4) to ++(3/8*\w,0) node[ocirc] (GH4){} to++(1/8*\w,0) |- (I2);

        \node[draw, BrickRed, rectangle, minimum width=15/8*\w, minimum height=3.25*\h, right=3.5/8*\w of A, align=center, text width=\w] (Z) {$\Y_\text{T1} = \Y_B + \Y_C$};
        \node[BrickRed, above=1pt of Z] {Tier 1};
        \node[draw, BrickRed, rectangle, minimum width=15/8*\w, minimum height=5.25*\h, right=2/8*\w of Z] (Y) {};
        \node[draw, BrickRed, rectangle, minimum width=15/8*\w, minimum height=3.25*\h, right=2/8*\w of Y, text width=\w, align=center] (X) {$\Y_\text{T3} = \Y_G + \Y_H$};
        \node[BrickRed, above of=F,yshift=2mm, text width=\w, align=center] {$\Y_\text{T2} = \Y_D+ \Y_E + \Y_F$}; 
        \node[BrickRed, above=1pt of Y] {Tier 2};
        \node[BrickRed, above=1pt of X] {Tier 3};

        \node[draw, agentcolor, rectangle, minimum width=63/8*\w, minimum height=6*\h, right=-3.4*\w of Z, yshift=2mm] (Y) {};
        \node[agentcolor, above left=1pt of Y, xshift=3cm] {Supply Chain};
        
        \node[agentcolor, above of=A,yshift=20mm, xshift=15mm] {$\T_{\text{SC}} = {\color{PineGreen} \T_{\text{OEM}} } \cdot {\color{BrickRed} \T_{\text{T1}} } \cdot  {\color{BrickRed} \T_{\text{T2}} } \cdot 
        {\color{BrickRed} \T_{\text{T3}} } $};

    \end{circuitikz}
    }
    \caption{Diamond-shaped supply chain network and the determination of its equivalent parameter model $\T_{\text{SC}}$.   }
    \label{fig:diamond}
\end{figure*}
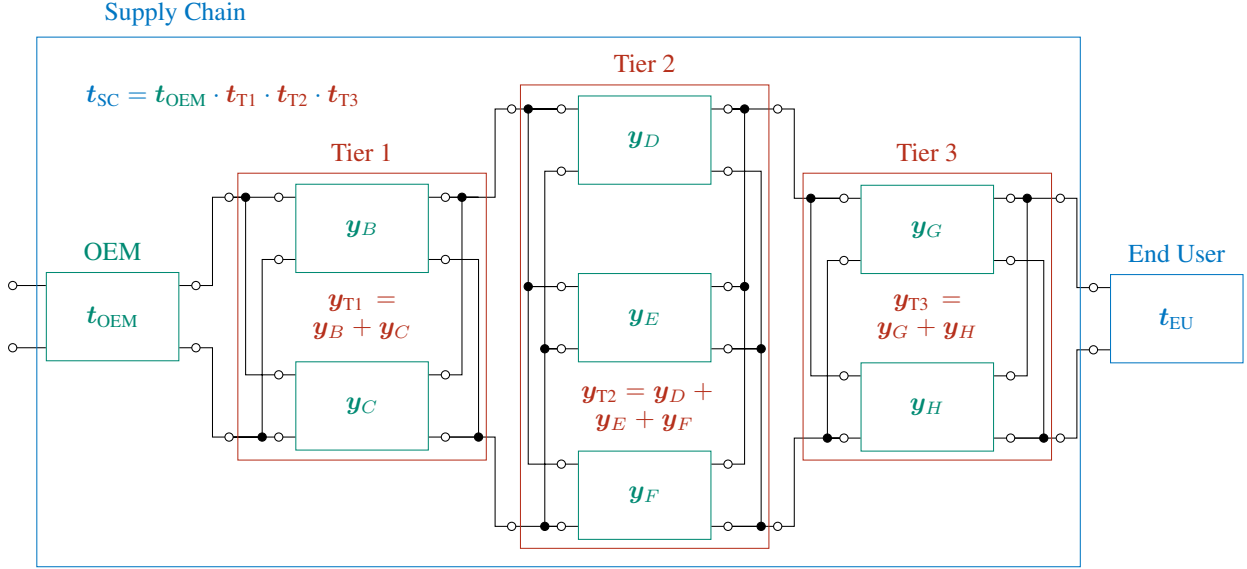

We illustrate the aggregation procedure with the supply chain in Figure \ref{fig:diamond}. 
In the supply-chain literature, this particular chain is known as a diamond-shaped supply chain  \cite{Slowinski2013}.  It consists of an original equipment manufacturer, or OEM, at the source, who delivers products through three intermediary tiers to an end user.  

The intermediaries within each tier compete, and we thus model them using $\Y$ parameters.  To obtain the equivalent parameter model for a tier, we sum the constituent matrices.  Then, to place them in the chain, they are converted to $\T$ parameter models.  For a generic tier's $\Y$ parameter, this is done as follows (see \cite{alexander2012fundamentals}):
\begin{equation}
    \T = \frac{1}{\Y_{12}} \begin{pmatrix}
        -\Y_{11} & 1\\
        \det(\Y) &-\Y_{22}
    \end{pmatrix} .
\end{equation}

Doing this for all three tiers leads to parameter matrices $\T_\text{T1}$, $\T_\text{T2}$, and $\T_\text{T3}$.  Finally, the $\T$-parameter matrix for the entire supply chain is found by multiplying these together with the OEM's parameter as follows: 
\begin{equation}
   \T_\text{SC} =  \T_\text{OEM} \cdot \T_\text{T1} \cdot \T_\text{T2} \cdot \T_\text{T3} .
\end{equation}

The chain's parameter matrix unambiguously determines the behavioral operator of the entire chain.  It does not matter that the individual intermediary or tiers have entirely dissimilar behavioral operators and, thus, are strongly heterogeneous.  If necessary, a representative tier or intermediary can be found using the methods outlined in the previous section.  The aggregate behavior of the chain, however, is not limited to a scaled-up version of a single, homogeneous agent, as \cite{caselli2000representative} has criticized the concept for.

\section{Applications}
\label{sec:Applications}

In this concluding section, we apply the theory to build four network models of increasing complexity in the following subsections: 
\begin{itemize}
\item[\ref{sec: robinson}] A Robinson Crusoe economy to show how a classic economic textbook problem is reformulated and rendered fully dynamic by the theory.
\item[\ref{sec:bullwhip}] A multi-level supply chain to show how a frequency domain analysis can provide insight into the bullwhip effect. 
\item[\ref{sec:Batteries}] An electricity market, where we add a battery storage system to show how the modularity of theory allows designers to build hypothetical scenarios for analysis of the resulting duck curves. 
\item[\ref{sec:MacroeconomicModel}] A complete network model of an economy, featuring multiple layers of abstraction down to the micro level.  We generate theoretical GDP, inflation, and interest rate forecasts. 
\end{itemize}

For each application, we design the system model in a top-down fashion.  At the top level of abstraction, we take the entire model as a single multiport, whose ports, if any, accommodate the exogenous inputs to it.   At the bottom level lies the component level where the elementary agents make up an economic circuit.  All models feature at least one intermediate level.   The economy, in particular, features four additional levels to show how a complex model can be built this way.  These include going from the top downward, a macro level with its sectors, then one with subsectors,  and a level with the individual agents, before finally reaching the micro level.  

To analyze the emergent dynamics, we work bottom-up. Starting at the micro level, we aggregate until we arrive at the level of abstraction where the signals of interest are intercommunicated. The emergent dynamical behavior is obtained by simulating the models on the LTSpice circuit simulator \cite{Engelhardt2007LTspice}.  Flows or incentives are readily measured by, respectively, placing ammeters or voltmeters at the relevant interconnection.  We interpret the dynamical behavior by tracing causal paths, consisting of dependent/independent signal links, that lead from an imposed shock to some response of interest.  

The use of existing network simulators for economic networks has proven to be highly efficient, both in the design and execution phases.  They have been developed to design and run simulations of highly complex electronic networks efficiently, and this does carry over to economic networks.  Using an ordinary mid-level laptop, the simulations are practically instantaneous except for the model of the entire economy, which completes within a few seconds.

\subsection{The Robinson Crusoe Economy}
\label{sec: robinson}
The Robinson Crusoe economy, see e.g. \cite{Varian2010}, is a highly simplified model used in economics to illustrate how the effects of scarcity, production, consumption, and the labor-leisure tradeoff on the economic equilibrium.     
In this section, we design such a multiport network model that includes these effects.  We design top down and focus on the dynamics of the progression to equilibrium.  

\begin{figure}[ht]
    \centering
    \begin{subfigure}[t]{0.45\textwidth}
        \centering
        \raisebox{1cm}{\includegraphics[width=\linewidth]{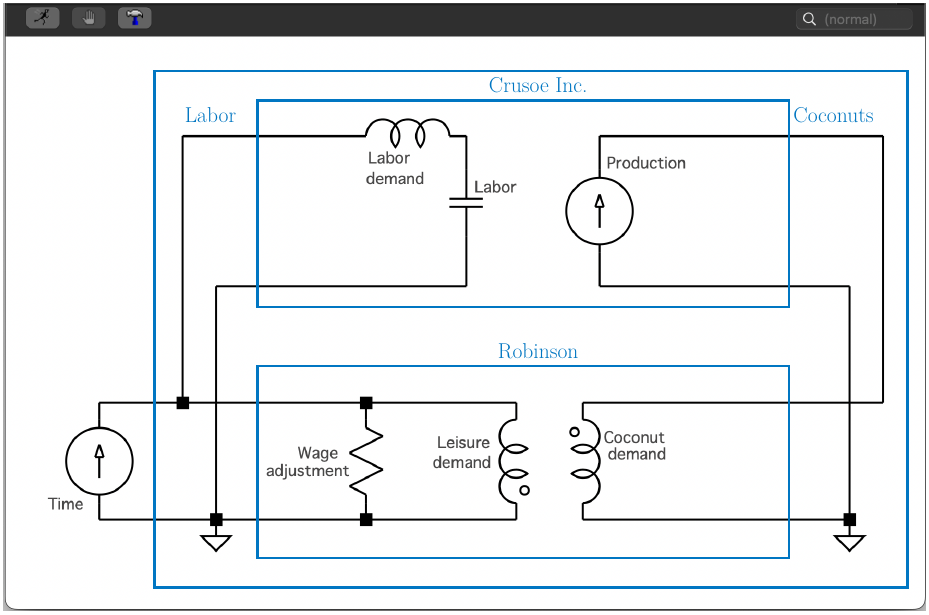}}
        \caption{Network Model}
        \label{fig:RobinsonStatic}
    \end{subfigure}%
    \hfill
    \begin{subfigure}[t]{0.45\textwidth}
        \centering
        \includegraphics[width=\linewidth]{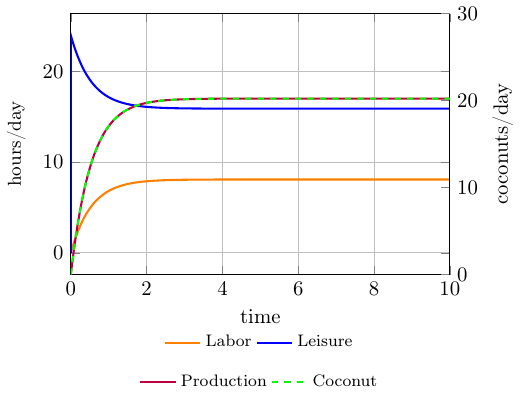}
        \caption{Simulation results}
        \label{fig:Robinsonplot}
    \end{subfigure}
    \caption{Simulating the Robinson Crusoe Economy as a multiport network in LTspice.}
    \label{fig:RobinsonCombined}
\end{figure}

A Robinson Crusoe economy features a single individual on a deserted island. There is no trade with the outside world, and only time is provided for.  This means that the whole economy is represented by a 1-port having a single port where time flows in at a constant rate. 
Figure \ref{fig:RobinsonStatic} shows an LTSpice implementation with the flow of time given by a current source.  

One abstraction level lower, Robinson Crusoe is split into a production and consumption sector called Robinson and Crusoe Inc., respectively.    Robinson supplies labor and demands coconuts, while Crusoe Inc. hires labor to produce those coconuts.  The sectors are thus both 2-ports, each with a port for labor and one for coconuts.  The interconnections given in the figure reflect that the sectors compete for the available time and that the economic surplus is divided between the two.   

Robinson's trade-off between leisure and coconuts is modeled as in the consumer example in Section \ref{sec:InteractionsWithPorts} and Figure \ref{fig:Consumer}.  Here, it models how Robinson alters his demand for leisure based on the price of coconuts and vice versa. The labor provided by Robinson to Crusoe Inc. is the available time less the leisure demand.

Crusoe Inc.'s production function is modeled as a simplified version of the producer example in Section \ref{sec:MarginalProduct} and Figure \ref{fig:Multivarprod}, featuring only the labor factor. 

We investigate the dynamics of the economy as it moves from its initial conditions to its equilibrium state.  In Figure~\ref {fig:Robinsonplot}, we plot the demand for leisure, labor, and coconuts in time coming from the LTSpice implementation.  The results are readily understood by analyzing the dynamics at the component level as follows: 

\begin{figure}
    \centering
    \includegraphics[width=0.7\linewidth]{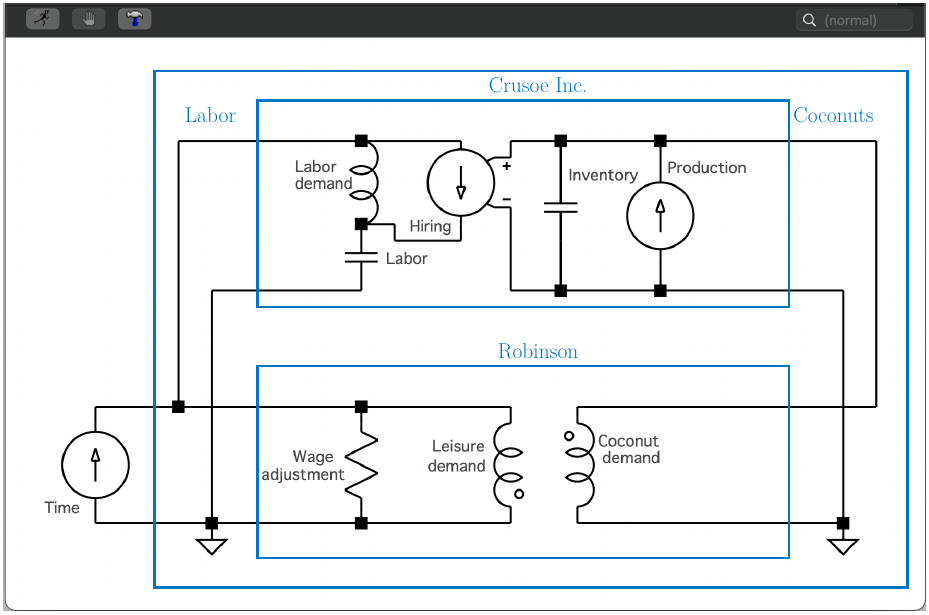}
    \caption{Robinson Crusoe Economy with extended Crusoe Inc. 2-port.}
    \label{fig:RobinsonDynamic}
\end{figure}

\begin{figure}[ht]
    \centering
    \begin{subfigure}[t]{0.48\textwidth}
        \centering
        \includegraphics[width=\linewidth]{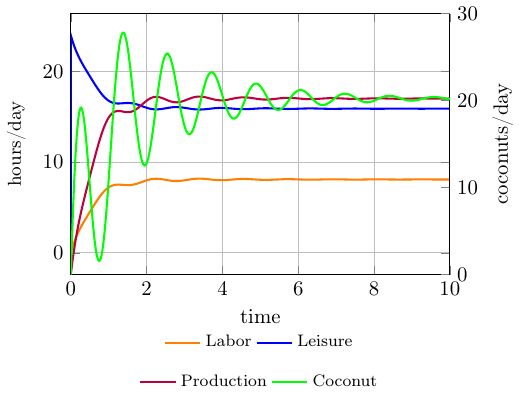}
        \caption{Effect of inventory on economic dynamics.}
        \label{fig:RobinsonInventory}
    \end{subfigure}%
    \hfill
    \begin{subfigure}[t]{0.48\textwidth}
        \centering
        \includegraphics[width=\linewidth]{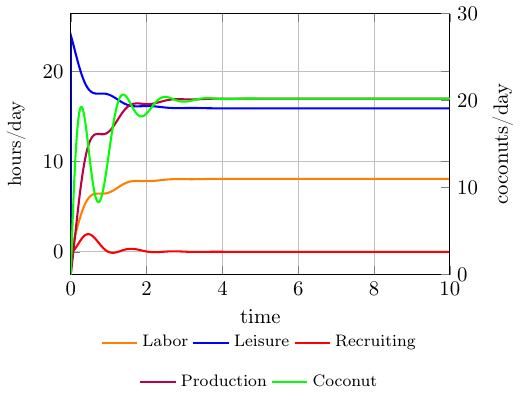}
        \caption{Active recruitment damps economic dynamics.}
        \label{fig:RobinsonRecruiting}
    \end{subfigure}
    \caption{Inventory and recruitment extensions in the Robinson Crusoe economy.}
    \label{fig:RobinsonExtensions}
\end{figure}

Before Robinson Crusoe's arrival, there was no demand for labor, leisure, or coconuts.  Upon arrival, Robinson initially places no value on his leisure time (blue) and 
flows it all through the shunt resistor to the ground at zero desirability. 
{Proportional to this flow, an} incentive is built up over the resistor to get to work. This incentive is communicated to Crusoe Inc., which hires Robinson to work, causing the demand for labor to ramp up (orange).  It also incentivizes Robinson to raise his price for leisure time and demand more of it. 
{
The labor hired by Crusoe Inc. is used to produce (red) the coconuts demanded by Robinson (green). The mutual inductance between Robinson's demand for leisure and coconuts determines the equilibrium state observed in the figure.}


We next complete the model by including the effects of scarcity on the dynamics.  This is done by adding a capacitor that represents a storage where coconuts are held in inventory (see Figure \ref{fig:RobinsonDynamic}).  When coconuts are scarce, inventories are low, and vice versa.   In Figure \ref{fig:RobinsonDynamic}, we plot the dynamics of the various demand flow quantities.  These exhibit second-order dynamics. The presence of an LC circuit in the coconut market results in strong seasonal fluctuations (green).  
Due to the cross-elasticities of demand for leisure and coconuts, the effect spreads out to the labor market while dampening out in the process due to wage adjustments.

In practice, to mitigate such fluctuations in demand, Crusoe Inc. might hire labor when inventories are low and dismiss labor when they are high.  In Figure \ref{fig:RobinsonDynamic}, we do this by including a controlled current source that actively regulates the demand for labor based on the inventory level. The results in Figure 
 \ref{fig:RobinsonRecruiting} shows that this significantly reduces the fluctuations.   By adjusting the sensitivity of the controlled source, Crusoe Inc.'s hiring policy can be tuned to yield a desired demand profile.

\subsection{Filtering Effect of Intermediaries in Supply Chains}
\label{sec:bullwhip}
In this example application, we analyze a supply chain in the frequency domain.   Supply chains often exhibit the bullwhip effect (see e.g. \cite{Ouyang2010}), where small fluctuations in supply or demand grow larger as they move downstream. Such effects are naturally analyzed in the frequency domain.  Here we show how the multiport framework allows us to study the chain as a cascade of signal processing filters and to use tools like Bode plots to analyze how signals are transmitted, delayed, or amplified across the supply chain.

The entire chain is modeled as a 1-port, whose port is facing upstream to an exogenous producer.  The chain has a lower level consisting of three tiers: a supplier, an intermediary, and a customer.  Both the supplier and intermediary are 2-ports, receiving the goods from upstream and delivering them downstream.  The customer is a 1-port and terminates the chain.   The multiports are interconnected in cascade.  The chain thus functions to divide the economic surplus by passing it along from producer to consumer.

\begin{figure}[ht]
    \centering
    \includegraphics[width=.95\linewidth]{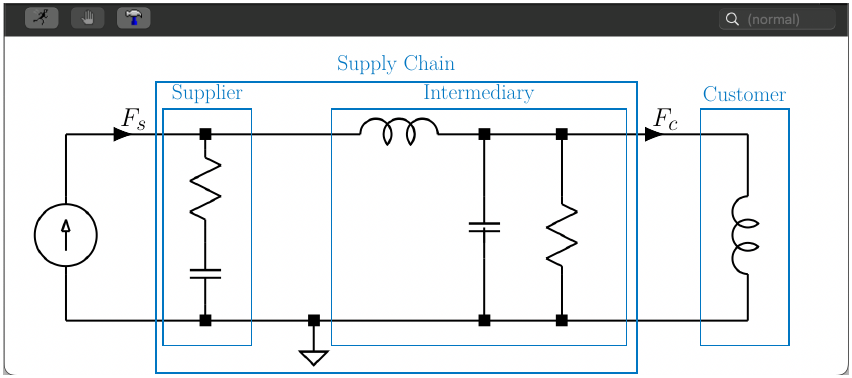}
    \caption{ Rudimentary supply chain as a multiport. }
    \label{fig:MatchingNetwork}
\end{figure}

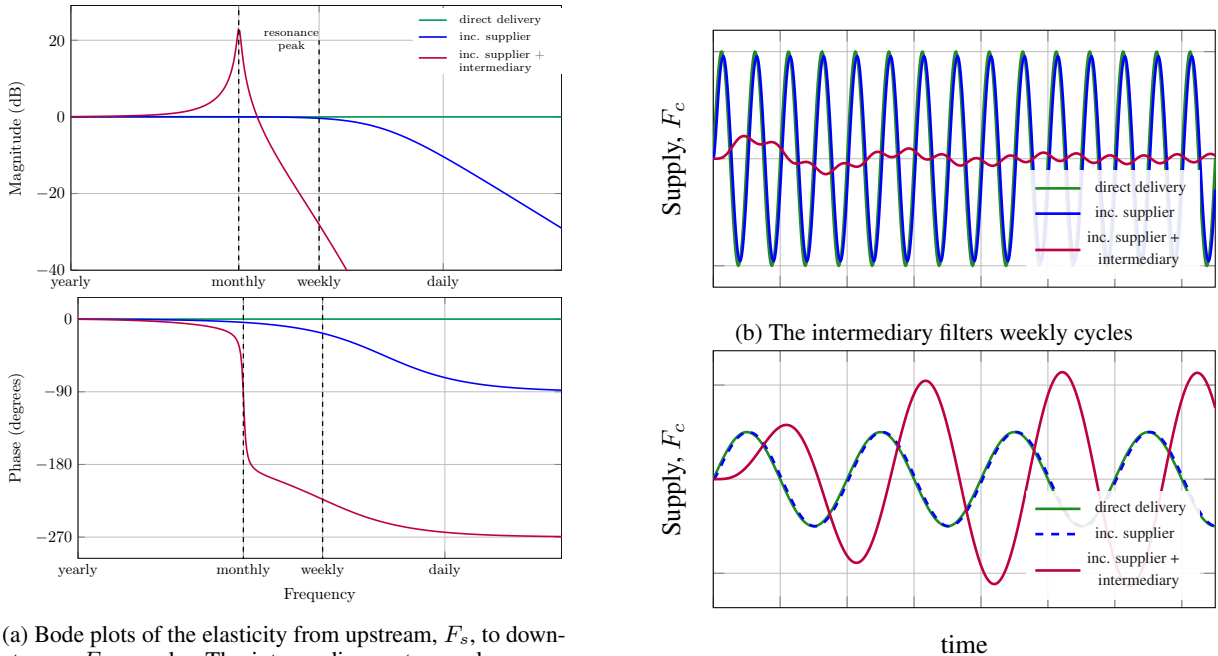
\begin{figure}[ht]
    \centering
    \begin{minipage}[h]{0.45\textwidth}
        \centering
        \begin{subfigure}{\linewidth}
        \centering
        \includegraphics[width=\linewidth]{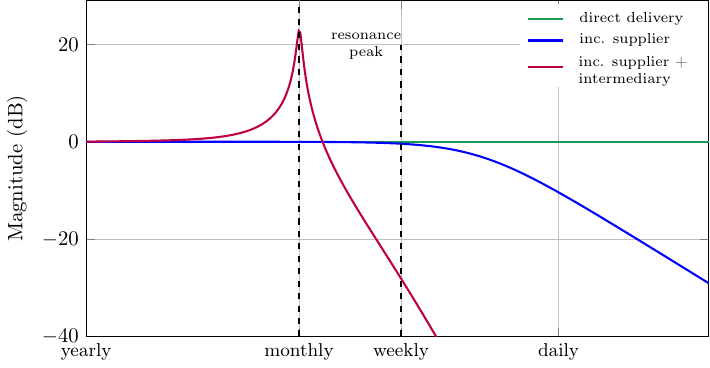}
        \includegraphics[width=\linewidth]{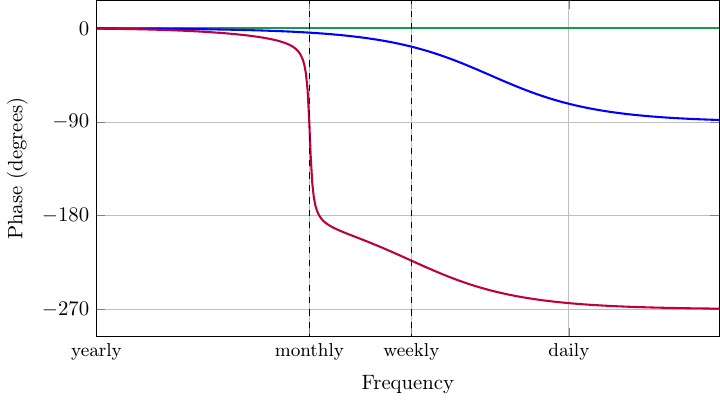}
        \caption{Bode plots of the elasticity from upstream, $F_s$, to downstream, $F_c$, supply. The intermediary acts as a low-pass filter, with a cut-off at monthly supply cycles. The resonance peak at this cycle is the cause of the bullwhip effect.}
        \label{fig:Bode}
         \end{subfigure}
    \end{minipage}%
    \hfill
    \begin{minipage}[]{0.5\textwidth}
        \centering
        \begin{subfigure}{\linewidth}
            \centering
\begin{tikzpicture}
  \begin{axis}[
    ylabel={Supply, $F_c$},
    grid=major,
    width=\linewidth,
    height=5cm,
    xmin=0,
    xmax=150,
    xticklabels = {},
    yticklabels = {},
    legend pos=south east,
    legend style={draw=none, fill=white, fill opacity=0.9,font=\tiny},
    legend cell align={left},
  ]
    \addplot[mark=none, ForestGreen, line width=1pt]
      table [x index=0, y index=1, col sep=space] {_main_matter/Multiport/filtered.txt};
    \addlegendentry{~~direct delivery};

    \addplot[mark=none, blue, line width=1pt]
      table [x index=0, y index=2, col sep=space] {_main_matter/Multiport/filtered.txt};
    \addlegendentry{~~inc. supplier};

    \addplot[mark=none, purple, line width=1pt]
      table [x index=0, y index=3, col sep=space] {_main_matter/Multiport/filtered.txt};
    \addlegendentry{\shortstack[r]{inc. supplier +\\intermediary}};
  \end{axis}
\end{tikzpicture}

            \caption{The intermediary filters weekly cycles}
            
        \end{subfigure}

        \begin{subfigure}{\linewidth}
            \centering
            \begin{tikzpicture}
                \begin{axis}[
                    xlabel={time},
                    ylabel={Supply, $F_c$},
                    grid=major,
                    width=\linewidth,
                    height=5cm,
                    xmin=0,
                    xmax=150,
                    xticklabels = {},
                    yticklabels = {},
                    legend pos=south east,
                    legend style={draw=none, fill=white, fill opacity=0.9, font=\tiny},
                    legend cell align={left},
                ]
                    \addplot[mark=none, ForestGreen, line width=1pt] table [x index=0, y index=1, col sep=space] {_main_matter/Multiport/Bullwhip.txt};
                    \addlegendentry{~~direct delivery};
                    \addplot[mark=none, blue, dashed, line width=1pt] table [x index=0, y index=2, col sep=space] {_main_matter/Multiport/Bullwhip.txt};
                    \addlegendentry{~~inc. supplier};
                    \addplot[mark=none, purple, line width=1pt] table [x index=0, y index=3, col sep=space] {_main_matter/Multiport/Bullwhip.txt};
                    \addlegendentry{\shortstack[r]{inc. supplier +\\intermediary}};
                \end{axis}
            \end{tikzpicture}
            \caption{The intermediary amplifies monthly cycles.}
        \label{fig:bullwhip}
        \end{subfigure}

    \end{minipage}
    \caption{Frequency and time-domain analysis of the effect of the intermediary in the supply chain.}
    \label{fig:intermediary}
\end{figure}

To investigate the dynamical behavior, consider first the supply chain without the supplier and the intermediary, i.e., direct delivery of the upstream supply to the customer. The producer is modeled as an AC current source, and the customer by an inductor that describes its demand schedule.  The Bode diagram in Figure \ref{fig:Bode} shows  0 dB amplification and 0-degree phase shift across the entire frequency spectrum (\textcolor{ForestGreen}{green}). This means that a change in upstream supply results in an immediate and unamplified change in downstream supply to the customer. 
In practice, this would correspond to a situation of volatile prices and inventories.

Next, we include the supplier. It has a capacitor for local storage and a resistor for carrying costs.  With it, we obtain the behavior given by the \textcolor{blue}{blue} plots.  The magnitude plot shows that the volatility in production on a weekly or higher frequency are filtered out, which means that the customer faces more stable prices and inventory levels.
The phase plot shows, in addition, that a delay of up to a quarter of the cycle period is to be expected for high-frequency production shocks.
The fluctuations in the flow to the customer will lag those in the flow from the external production.

Last, we include an intermediary.  It is modeled conform the trader example in Section \ref{sec:TraderAsRLCCircuit} and Figure \ref{fig:trader}.  The corresponding Bode plots are given in \textcolor{purple}{red} and show the profile of a second-order system. Below the monthly rate, the fluctuations are mostly passed along to the customer. At the monthly rate, there is a sharp resonance peak at which they are strongly amplified, and deliveries become suddenly delayed.  
At higher frequencies, the response of the chain rapidly falls off, with delays increasing gradually.

The resonance peak is the rate at which we should expect effects like the bullwhip to occur.  Its location is mostly determined by the intermediary's storage elasticity as given by the capacitor elastance.  Higher elasticities correspond to more aggressive inventory management, and this shifts the resonance peak to the right.   This makes the bull whip less likely to occur.  However, this may subject the customer to undesirable high-frequency noise as the chain filters out a smaller band of high frequencies.   

The intensity of the bullwhip effect is given by the height of the resonance peak.  This is determined by the damping ratio \cite{franklin2019feedback}.  By adjusting its price, storage, and friction elasticities, the intermediary can decrease the peak and help to mitigate the effect.  

The general technique for achieving these desiderata is known as loop shaping in control theory \cite{franklin2019feedback}.  With it, supply chains can be tuned to track the reference output rate given by the producer, while rejecting unwanted disturbances and attenuating any spurious noise.  In this way, the chain provides customers with a smooth, shock-free supply while minimizing returns to the producer.

In Figure \ref{fig:intermediary}, the time-domain plots illustrate how the intermediary alters the propagation of cyclical supply variations. In panel (b), representing a weekly input cycle, the downstream supply (\textcolor{purple}{red}) closely tracks the upstream supply (\textcolor{blue}{blue}) with only a slight lag and reduced amplitude, indicating that high-frequency fluctuations are effectively filtered. In panel (c), corresponding to a monthly input cycle near the resonance frequency, the downstream response is not only delayed but also amplified relative to the upstream signal—a hallmark of the bullwhip effect. The phase lag visible in both cases is consistent with the negative phase shifts in the Bode plot, but the magnitude and sign of the response differ sharply depending on the driving frequency.

It is straightforward to extend this analysis to multi-tiered supply chains with multiple intermediaries connected in cascade. In such cases, the overall Bode plot of the chain is obtained by summing the Bode plots, i.e., both the magnitude and phase plots, of the individual tiers.  

The procedure allows us to understand the bullwhip effect.  If the tiers are not too dissimilar in their dynamic behavior, the location of the resonance peak will not differ too much. In this case, if a tier gets excited at its resonance frequency, the subsequent tiers will further amplify the fluctuations as they pass through the chain. 

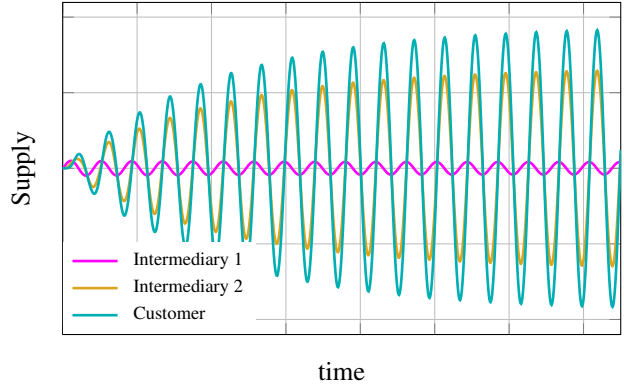
\begin{figure}
   \centering
       \begin{tikzpicture}
            \begin{axis}[
                xlabel={time},
                ylabel={Supply},
                grid=major,
                width=9cm,
                height=6cm,
                xmin=0,
                xmax=150,
                xticklabels = {},
                yticklabels = {},
                legend pos=south west, 
            legend style={draw=none, fill=white, anchor=south west, at={(0,0)}},
            legend style={font=\scriptsize}, 
        legend cell align={left},
                    ]
    
                \addplot[name path=demand, mark=none, Magenta, line width=1pt] table [x index=0, y index=3, col sep=space] {_main_matter/Multiport/Bullwhip2.txt};
                \addlegendentry{~~Intermediary 1};
                \addplot[name path=demand, mark=none, Goldenrod, line width=1pt] table [x index=0, y index=2, col sep=space] {_main_matter/Multiport/Bullwhip2.txt};
                \addlegendentry{~~Intermediary 2};
                \addplot[name path=demand, mark=none, TealBlue, line width=1pt] table [x index=0, y index=1, col sep=space] {_main_matter/Multiport/Bullwhip2.txt};
                \addlegendentry{~~Customer};
        \end{axis}
        \end{tikzpicture}   
   \caption{Bullwhip effect: seasonality amplifies across the supply chain.  }
\label{fig:bw}
\end{figure}

In Figure \ref{fig:bw}, we show the measurements of the flow through each tier in the LTSpice simulation of a three-tiered supply chain.  This chain extends the chain in Figure \ref{fig:MatchingNetwork} by inserting an identical intermediary as an additional tier.  The bullwhip effect is evident from the increase in the amplitude of the fluctuations as we go down the tiers.   Interestingly, the analysis uncovers a related bullwhip effect on the time lags.  This is evident from the accompanying phase shift of the signals.  The entire bullwhip effect, thus, occurs in both magnitude and phase.  

\subsection{The Impact of Battery Energy Storage Systems on Electricity Markets}
\label{sec:Batteries}

\begin{figure}
    \centering
\begin{subfigure}{.48\linewidth}
    \includegraphics[width=\linewidth]{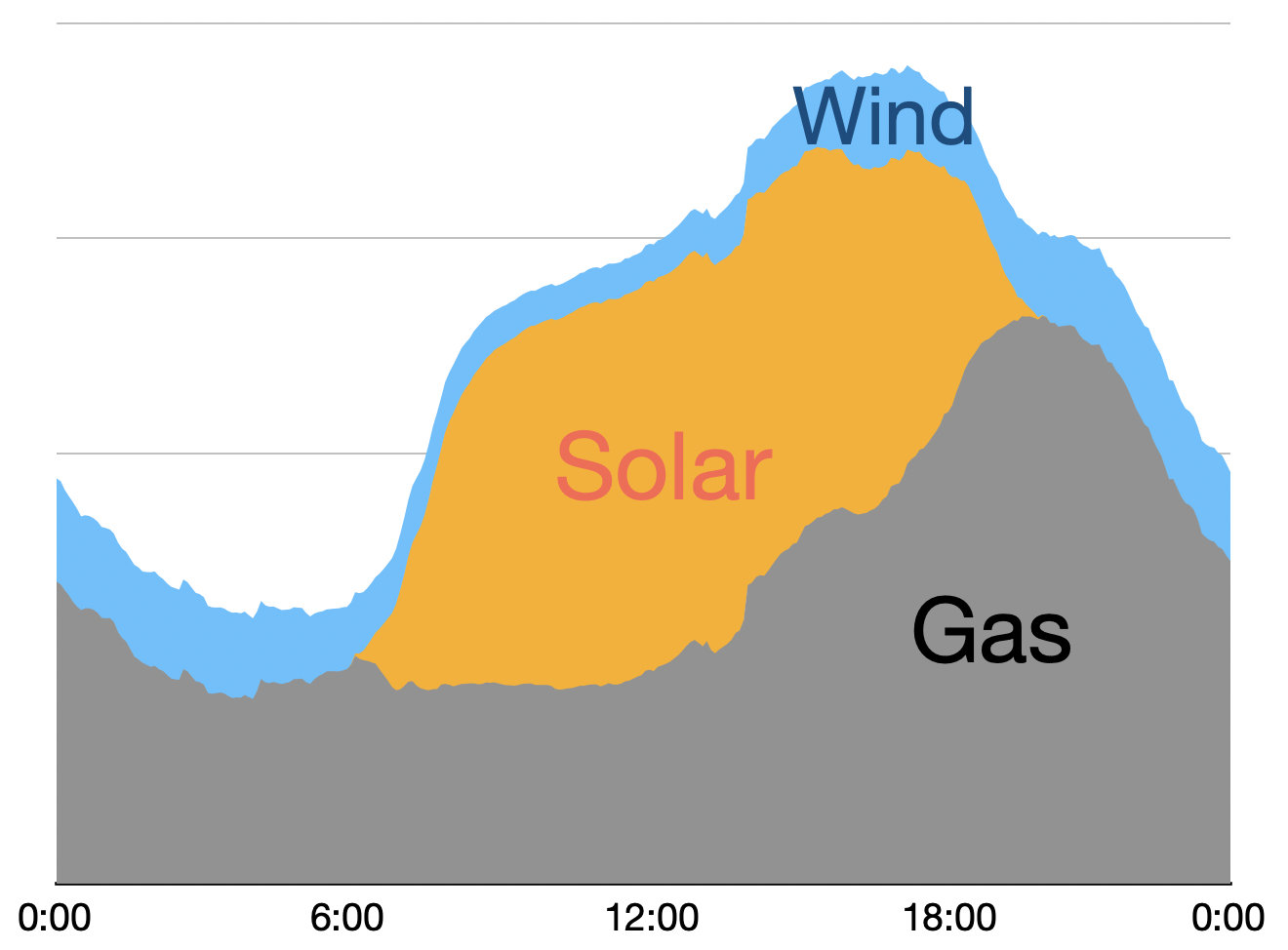}
    \caption{}
    \label{fig:CAISO2020}
\end{subfigure}    
\hfill%
    \begin{subfigure}{.48\linewidth}
    \includegraphics[width=\linewidth]{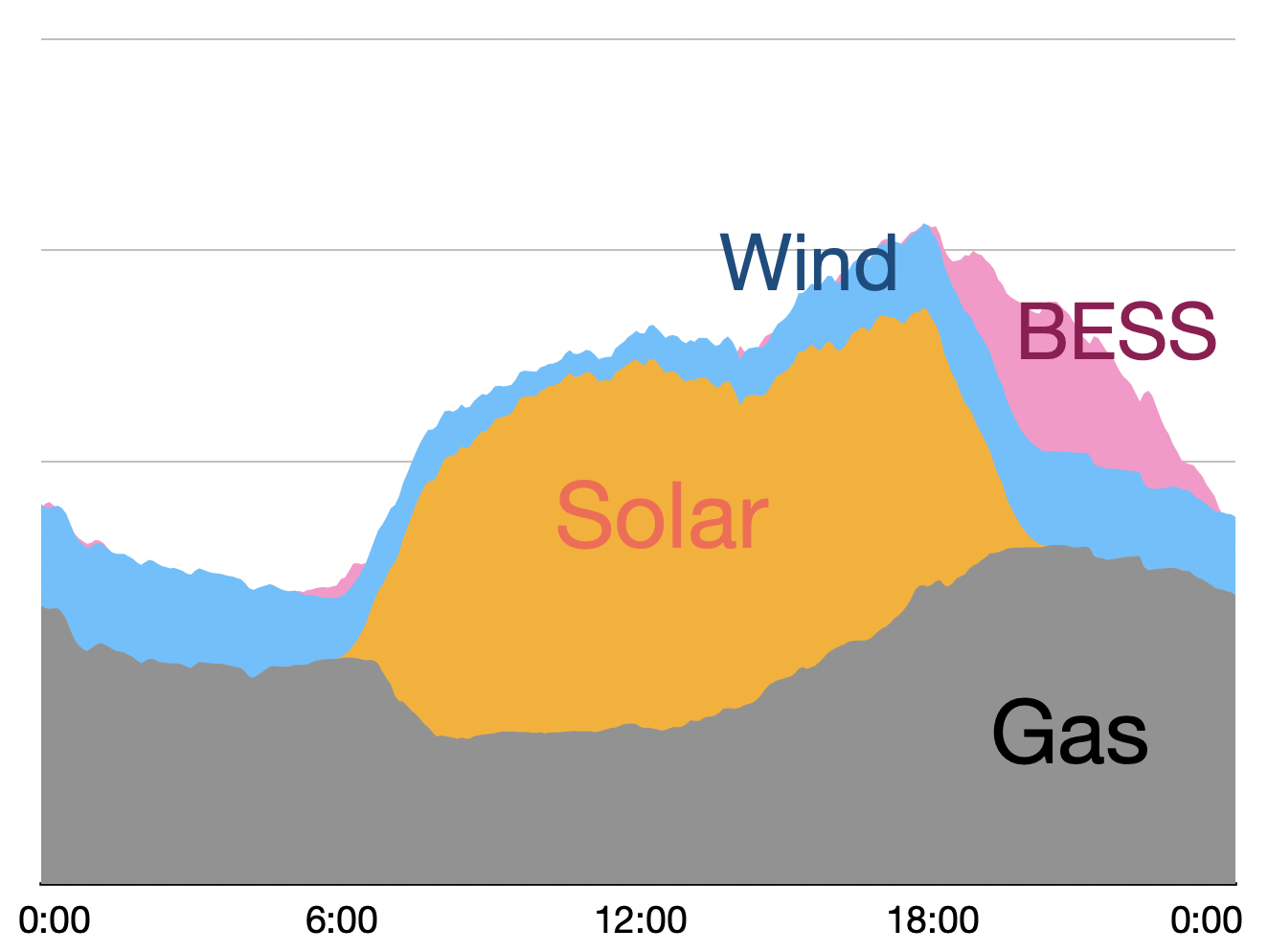}
    \caption{}
    \label{fig:CAISO2025}
\end{subfigure}    
    \caption{Electricity production per selected energy source in California on July 17, 2020 (a) and July 17, 2024 (b). Data from \cite{gridstatus2025}.}
    \label{fig:CAISO}
\end{figure}

Contemporary electricity markets are becoming more complex due to the integration of renewable energy sources and rapid technological advancements (see, e.g. \cite{erdiwansyah2021}). To examine supply, curves are drawn that give the electricity generated by each source at each moment in the day. 
These curves show a common problem in energy systems with high solar penetration: Because of a timing imbalance between peak solar production at midday and peak demand during the evening hours, renewable energy sources need to be curtailed while fossil generators need to step in to meet peak demand.
This is borne out by the duck-like shape \cite{Agmad2020} of the graph for natural gas (\textcolor{DarkGray}{gray}) in Figure \ref{fig:CAISO2020}.

To increase the market's efficiency and reduce the carbon footprint, energy storage systems are added.  With these, excess renewable production can be shifted to the times when demand peaks.  Figure \ref{fig:CAISO2025} shows the electricity mix\footnote{For the purpose of presentation we have excluded data on electricity sources other than natural gas, wind, solar, and batteries.} in California on the same day in 2024, after such systems have been added. Excess solar production is now stored during the day and made available during peak demand, reducing the natural gas production during the evening.

However, rather than analyzing post hoc, it is advantageous to predict what, in general, the effects are of adding actors to the market.  Electricity supply per source can be {modeled}  analytically with multiport networks. They, therefore, present an opportunity to fulfill that purpose. 

\begin{figure}
    \centering
    \includegraphics[width=.7\linewidth]{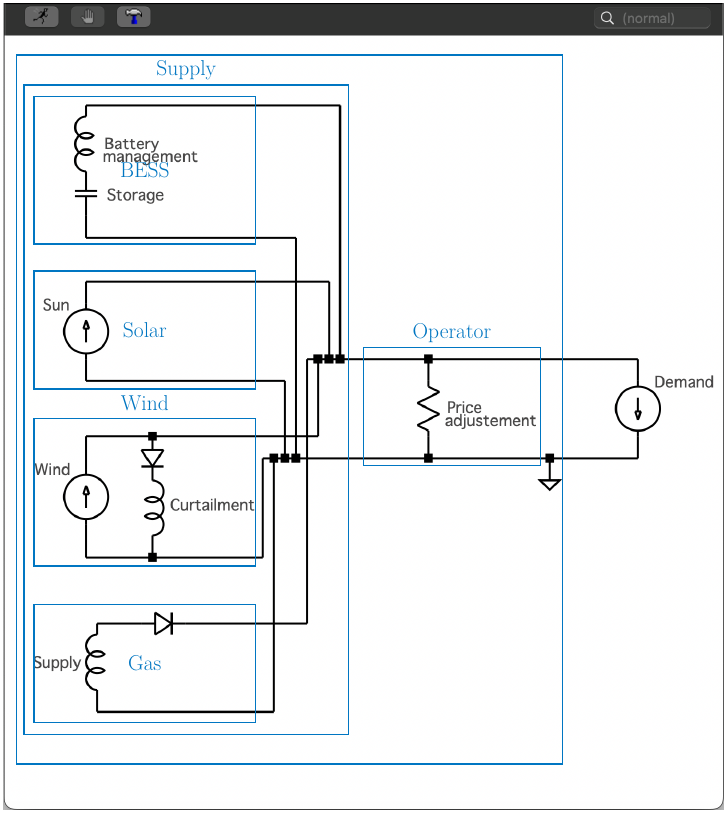}
    \caption{Electricity-market multiport model in LTspice. }
    \label{fig:SPICEElectricity}
\end{figure}

Figure \ref{fig:SPICEElectricity} contains our multiport network model as it appears in LTSpice. We model the supply market as a single 1-port,  on which the demand is exogenously imposed at its port.   Demand, which we assume to be inelastic, is modeled by a current source that is programmed to follow the typical pattern of electricity demand during a day.


At a lower level of abstraction, the market is split into an energy generation sector and a system operator that transmits the energy from the generators to the exogenous demand.  Proceeding further downward, the generator 1-port consists of four individual generators interconnected to be in competition with each other: gas, wind, solar, and a Battery Energy Storage System (BESS).   Total production appears at the port, where it is transmitted to the supply-facing port of the operator. The latter is a 2-port which, at its demand-facing port, distributes the produced electricity to the consumers.   
 
\begin{figure}
    \centering

\begin{subfigure}{.48\linewidth}
    \centering
    \includegraphics[width=\textwidth]{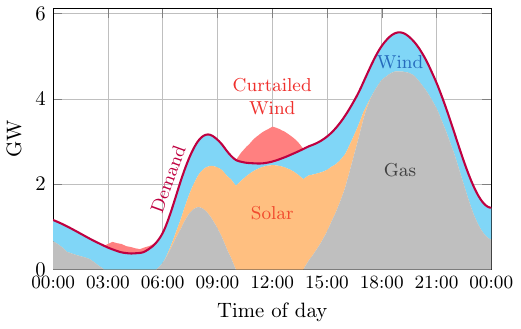}
    \caption{Without BESS: Wind is curtailed when demand is low and gas fulfills most of peak demand.}
    \label{fig:Curtailment}
\end{subfigure}
\newcommand{\minexpr}[2]{\pgfmathparse{min(#1,#2)}\pgfmathresult}
\begin{subfigure}{.48\linewidth}
    \centering
    \includegraphics[width=\textwidth]{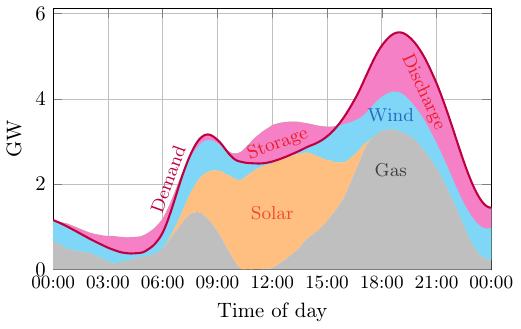}
    \caption{With BESS: Excess wind and solar electricity is stored by the BESS and discharged during peak demand.}
    \label{fig:Storage}
\end{subfigure}
\caption{Duck curve simulations.}
\label{}
\end{figure}

We first consider the market without a BESS.   Figure \ref{fig:Curtailment} displays the electricity production per source during a particular day generated by the network. Notice that gas-powered production takes on the distinct duck-like shape seen in empirically determined curves, as in Figure \ref{fig:CAISO2020}.  Our analytical solution, however, puts us in a position to investigate the causes.  

To analyze this dynamic behavior, we examine the internal circuitry of the agents that determines the underlying dynamic behavior. The current source for solar energy is programmed to follow the available sunlight during the day, showing the distinctive peak at midday (\textcolor{orange!50}{orange}).  The correct wind curtailment (\textcolor{red!50}{red}) is achieved by shunting its current source with an inductor and a diode in series.  This ensures that no wind (\textcolor{cyan!50}{blue}) is delivered when the price on the inductor drops below zero.  The inductor in gas makes it purely price-driven, with the diode ensuring that it engages in production only.  The curve shows (\textcolor{gray!80}{gray}) how gas ramps up sharply in the evening when prices are high.

As a dynamical system, the network is almost trivially simple, having only two real-valued poles.  Yet, it is able to give a surprisingly accurate reproduction of comparable empirical curves.

Coming back to our goal, we now add a BESS to the supply sector for comparison.  It is modeled as an LC circuit.  The capacitor represents the battery where the energy is stored.  The inductor records its internal reservation price and provides the appropriate incentives to store or discharge.  This adds two complex-conjugate poles, making the model fourth order.  

The behavior of the resulting electricity market is shown in Figure \ref{fig:Storage}.  Comparing it with that in Figure \ref{fig:Curtailment}, we see that the wind curtailment bump at midday has disappeared, and gas production during peak demand is lowered. This effect is also corroborated by the empirically determined curve in Figure \ref{fig:CAISO2025}.  
The excess renewable energy during the afternoon hours is stored by the BESS (\textcolor{magenta!50}{pink}).   At peak demand in the evening, the BESS discharges, partly replacing the gas-powered production and thereby avoiding emissions.

\subsection{A Multiport Network Model of the Economy}
\label{sec:MacroeconomicModel}
\subsubsection{Introduction}
In this final application, we design and analyze the behavior of relatively complex model of an entire economy.  
The underlying network is taken from \cite{Berend}, to which we refer the reader for the details.  This concerns a rather comprehensive model that combines the circular flow, stock-flow consistent, ISLM, and Solow growth models from macroeconomic texts, together with a model of a commercial bank based on accounting principles.  To it, we have added a model for the security markets from \cite{Jeroen} and the diamond-shaped supply chain presented in the preceding.  

\subsubsection{The Model}
\label{sec:TheModel}
\begin{figure*}[ht!]
    \centering
    \includegraphics[width=\textwidth]{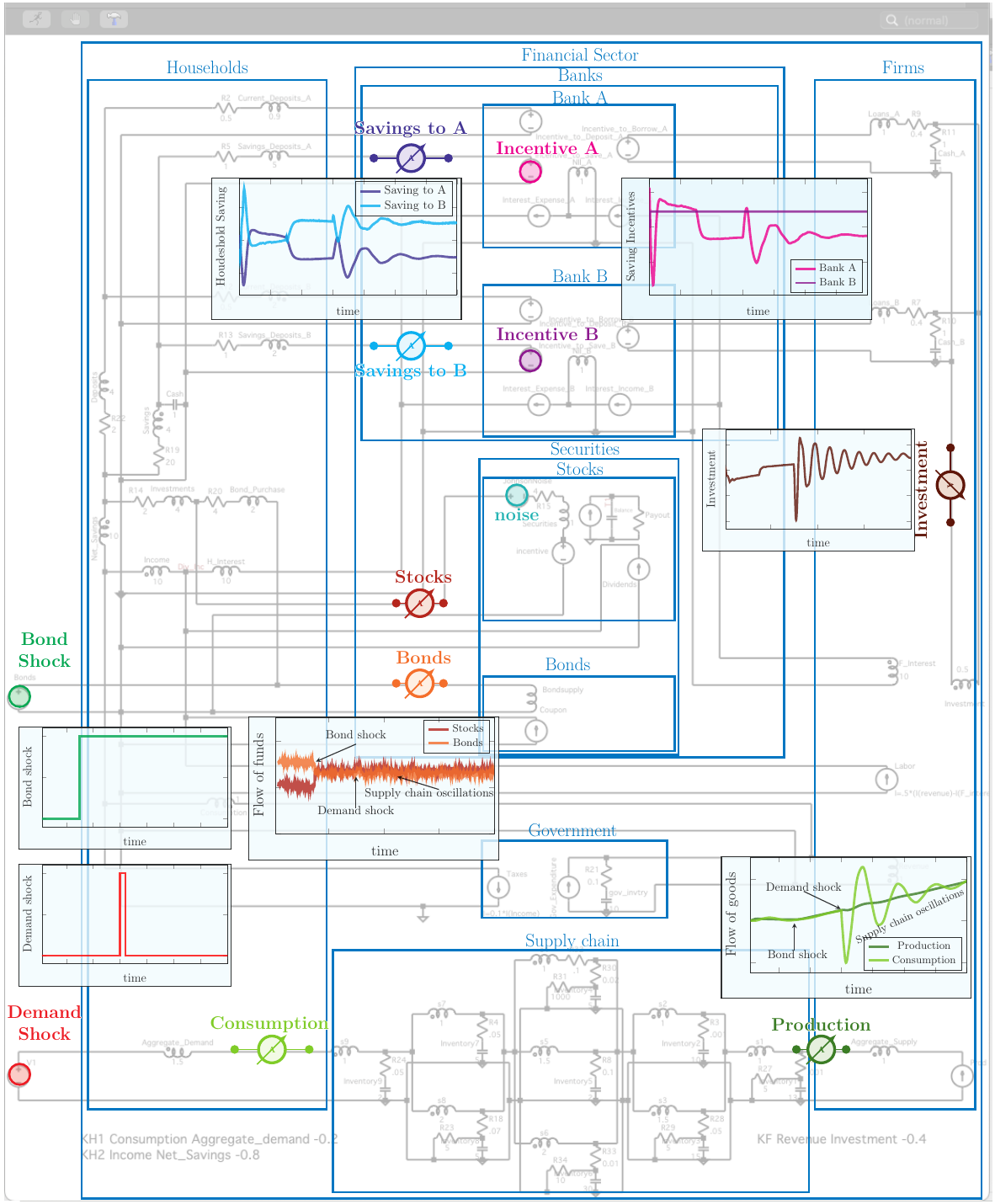}
    \caption{Macroeconomic multiport model in LTspice; plot colors match ammeter and source colors.}
    \label{fig:Macromodelwplots}
\end{figure*}

The final LTSpice layout is displayed in  Figure \ref{fig:Macromodelwplots}, in the background of the plots of the flows measured by the corresponding amp meters.  We have indicated these these by a circular icon marked with an A.  We have color-coded them to correspond to the graph of the corresponding signals that overlay the network.

We have represented the economy itself as a 2-port.  There is a port for the flow of foreign capital to the securities market (\textcolor{Green}{Green}), and one for the flow of goods to the productive sector (\textcolor{Red}{Red}).  Otherwise, the economy is closed.

At one abstraction level lower, the network contains five multiports.  This includes the sectors typical of a macro model: household, financial, firms, and government.  To this, we have added a supply chain as a demonstration of the modularity of the theory.  

We have modeled the financial sector to have two further levels of abstraction.  At one level lower, it consists of a banking sector and a sector representing the securities industry.  We have designed it to be heterogeneous, as we show below.  If needed, the methods in Section \ref{sec:AggregationAndTheOperatorAlgebra} can be exploited to 
determine the appropriate representative agents for the sector and its subsectors.   Here we focus on the LTSpice simulations.  
 
Further down the level, the banking sector contains two 5-port banks.  They are in competition with one another for assets and liabilities by offering interest payments on the former and charging interest on the latter.  The assets are loans to firms, and the liabilities consist of both household savings and current account deposits.  They have distinctly dissimilar policies:    
Bank A is an aggressive discounter, while Bank B offers households the traditional banking experience.  
This is borne out by the graph containing the incentive that each bank provides the households to save: Bank A rapidly adjusts this based on circumstances (\textcolor{Magenta}{Magenta}), while Bank B holds it steady (\textcolor{Plum}{Plum}). The circuit for this follows the example in Section \ref{sec:ReserveBankAsControlledSource} and Figure \ref{fig:bank}.   

The securities industry consists of a stock market and a bond market. Both are 2 ports. At one port, they compete for household investment. At the other port, they offer dividends and coupon payments, respectively, in return.  

The stock market adds significant volatility to its returns.  Although it is configured with a noise source (\textcolor{TealBlue}{TealBlue}), in reality, this is done in an entirely endogenous manner; see \cite{Jeroen} for the details.  The price volatility causes the flow of funds to also become volatile, both to the stock market (\textcolor{BrickRed}{BrickRed)} and somewhat less to the bond market (\textcolor{Orange}{Orange}).  At the flows of savings (\textcolor{BlueViolet}{BlueViolet} and \textcolor{ProcessBlue}{ProcessBlue}), it is still noticeable, and also Bank A's policy is slightly affected.  Beyond that, at the firms' capital investment (\textcolor{Sepia}{Sepia}), the rate of production (\textcolor{OliveGreen}{OliveGreen}), and demand (\textcolor{LimeGreen}{LimeGreen}), the effect ceases to be visible.  

The bond market, in contrast, provides a stable return on investment.  It maintains a current price for bonds at the inductor.  The current source makes the coupon payments based on the total face value of the bonds issued.  

An external port is connected to domestic investment to accommodate foreign investment.  After 1.5 years, the appetite of foreign investors to purchase bonds suddenly steps up (\textcolor{Green}{Green}).  As a result, the domestic flow of bonds steps down (\textcolor{Orange}{Orange}), and is redirected to the stock market (\textcolor{BrickRed}{BrickRed}).   

The shock then further propagates through the network.  Bank A, the aggressive discounter, reacts to the increase in savings and makes a downward revision to its savings incentive (\textcolor{Magenta}{Magenta}).  
Because Bank B keeps its incentive at the same level, household savings move away from Bank A (\textcolor{BlueViolet}{BlueViolet}) and take their business to Bank B (\textcolor{ProcessBlue}{ProcessBlue})
  The increase in total savings allows the banking sector to increase the size of its loans to firms, causing capital investment rates to show a relatively small increase (\textcolor{Sepia}{Sepia}).  This, subsequently, translates into an increase in production on the market for goods and services. There, household demand is incentivized to increase to keep the market in balance.  

Theoretically, this change then further reverberates through the economy, as the change in demand influences consumer expenditure and so on.  However, it is not productive to keep follow this, as the output from the LTSpice simulation contains the combined effect.


We model the government as a 2-port without any further lower levels of abstraction. It taxes household income and adds government expenditure to the firms.


Going beyond the typical textbook macro model, we add a separate market for goods and services.  Households demand goods and services, and firms produce them, and delivery is done through a supply chain.   

The price coupling between the goods and services versus the various flows of money is achieved through mutual inductances, comparable to the substitution effect in the example in Section \ref{sec:ConsumerWithSubstituteGoods}.  The circuitry is designed to operate in conformance with the IS-LM model\footnote{The acronym stands for Investment Savings vs Loanable Funds and Money; see \cite{Mankiw2016} }
to account for the various money flows and interest rates provided by the banking sector.  We refer the reader to \cite{Berend} for the details.

The supply chain is a 2-port system, with one port interconnected with the households and the other with the firms. Like the supply chain in Figure \ref{fig:diamond}, it is of diamond shape.  The intermediaries are modeled as RLC circuits, identical to the trader in Figure \ref{fig:trader}.  Each contributes two complex poles and, hence, the chain adds significantly to the complexity of its dynamic behavior.

This is evidenced by the economy's response to the shock in demand. At year 3, foreign demand shows a short impulse in its incentive to acquire goods.
Its suddenness causes massive imbalances in the supply chain, causing prices to drop in reaction to the inventory buildup. The drop in price causes demand to return, and the cycle repeats itself somewhat less markedly, leading to the damped oscillatory motion towards equilibrium characteristic of stable higher-order systems.  

The fluctuations then propagate through the entire economy. 
Households, when spending less on consumption, increase the rate at which they save (\textcolor{BlueViolet}{BlueViolet} and \textcolor{ProcessBlue}{ProcessBlue}) and invest (\textcolor{Orange}{Orange} and \textcolor{BrickRed}{BrickRed}), and vice versa.  Bank A immediately takes advantage of this, adjusts its client rates (\textcolor{Magenta}{Magenta}), and increases its share of total savings, only to reverse course when households increase consumption (\textcolor{LimeGreen}{LimeGreen}) again.  They thus follow the oscillations, lagging with a relatively small delay in phase.  A similar effect, albeit less noticeable,  is seen in the flow of funds to the securities market.   
Firms reduce capital investment (\textcolor{Sepia}{Sepia}) when production (\textcolor{OliveGreen}{OliveGreen}) is low and vice versa.  Fluctuating production, finally,  leads to oscillations in capital investment.  

Due to their interconnectedness,  households and firms additionally influence one another's responses.  The actual responses, as provided by the LTSpice simulation, are therefore the results of a highly complex set of interactions and communications between agents with all manner of internal feedback loops. The fact that the simulation runs within seconds is a testament to the potential of circuit theory for modeling the dynamics of complex economic systems.

\subsubsection{Forecasting and Analysis}
\label{sec:ForecastingAndAnalysis}
We conclude by showing how the foregoing model can be used to measure the typical variables of interest to economists. With their DSGE model in \cite{frbny_dsge_model} (reproduced here in Figure \ref{fig:DSGE}), the authors focus on forecasting GDP growth, inflation, and the rate of interest.  For the network model, these forecasts are determined simply by measuring the relevant flows in the LTSpice simulation.  

\begin{figure}
    \centering

        \centering
        \includegraphics[width=0.6\linewidth]{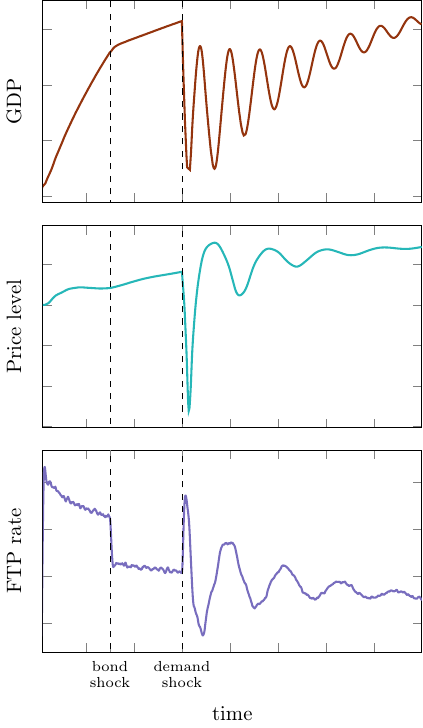}
\caption{Measurements of the GDP, the price level, and the FTP rate. }
\label{fig:Responses}
\end{figure}

\paragraph*{GDP}
To determine GDP, economists use the national income identity, which, for an open economy, reads $Y = C + I + G - \textit{NX}$. $Y$ is the flow of money that represents GDP.  The constituent flows are readily measured at the appropriate ports.  Consumption expenditure $C$ and investment $I$ at the households, government expenditure $G$ at its port, and net exports $\textit{NX}$ at the economy's external port.   Summing these measurements gives the projection of GDP shown in Figure \ref{fig:Responses}.   

At initiation, GDP grows steadily, but the rate slows down after the bond investment shock.  This is because the resulting increase in household investment in stocks decreases consumption expenditure $C$.  In addition, the accompanying decrease in tax revenue reduces government spending $G$.   These effects outweigh the increase in capital investment $I$.  

The effects of the demand shock on GDP are significantly more dramatic.
Immediately after the shock, GDP plummets due to the sharp decrease in consumption.  Due to the momentary nature of the shock, growth recovers on average after this.  However, due to the massive supply chain disruptions, GDP oscillates around its mean value, initially wildly so, which gradually dampens out to resume the previous steady growth rate.  At the onset of the COVID-19 pandemic, a similar demand shock took place, and its observed effects on the economy are comparable to these results.

\paragraph*{Inflation}

The addition of a separate goods market to the network allows the network to determine the price level and the inflation rate.  Economists typically conflate physical output with GDP, but this is unnecessary here.   
To measure the price level, we pair the flow of goods with that of the corresponding money. For retail price levels, we measure the ratio of household expenditure to the flow of goods to the households.   For wholesale price levels, we take the revenue against the flow to the firms.  Inflation is determined by taking the time derivative of the price levels. To obtain the rate of inflation, we take the logarithmic derivative instead.   Higher-order derivatives give the dynamics of inflation.

Figure \ref{fig:Responses} shows the retail price level, as measured in the simulation. Shortly after its initiation,  inflation disappears after a brief rise.  Then, due to the sudden availability of foreign funds following the bond investment shock, the inflation rate ticks up and remains steady until the onset of the demand shock.  The sudden drop in demand that follows has a strong deflationary effect, with prices falling rapidly.   As with GDP, inflation ultimately goes back to its previous level.  Supply chain dynamics do cause significant volatility, albeit relatively less pronounced than GDP.

\paragraph*{{Interest Rate}}
For the prevailing rate of interest, we use the banking sector's Funds Transfer Pricing or FTP Rate.  This is the average of the rate offered on savings and that charged on loans.  A more complete model might include a central bank within the sector that influences rates.  

Figure \ref{fig:Responses} contains the network's forecast for the FTP rate.  The response reflects the heterogeneity of the sector.  The overall downward trend is due to Bank B's intransigence on its savings rate, while being forced to compete for the issue of loans to the firms, thus bringing down the overall FTP rate.   Bank A's aggressive policies allow it to adjust both its client rates to closely follow the overall state of the economy.  Because of its actions, the interest rate serves to counteract the changes in the price level. Comparing the FTP rate with the inflation rate in Figure \ref{fig:Responses}, we see that the former is $180^\circ$ out of phase with the latter, thus acting countercyclically. 

Unlike GDP and the price level, the volatility in stock prices does influence the FTP rate.  This is particularly noticeable when the rate is relatively constant.  The banking sector competes for funds with the securities industry, and Bank A's aggressive policies subject the sector to any noise emanating from the stock market.

\subsubsection{Comparison with a DSGE Model}
We conclude by comparing the network with the DSGE model of the Federal Reserve Bank of New York \cite{frbny_dsge_model} (see Figure \ref{fig:DSGE}).  In their publication, the authors likewise concentrate on forecasting GDP growth, inflation, and the interest rate.  We focus on the individual letter combinations in the DSGE acronym.  

We first turn to the DS of Dynamic Stochastic and consider the fluctuations that appear in the forecasts.  In the DSGE model,  they are directly attributed to the stochastic shocks and are considered to be \textit{exogenous} (see, further, \cite{romer2016trouble}).  In the network, they are mainly caused by the supply chain and, hence, are \textit{endogenously} generated.  In fact, they point to the presence of second and higher-order dynamics in the chain (see Section \ref{sec:bullwhip}).  The shocks merely serve to excite the various frequencies:
The Brownian noise on the stock market is responsible for exciting the high-frequency fluctuations on the FTP rate and the demand shock the lower frequency oscillations in the various forecasts.  The latter shock, like the bond investment shock, is not even required to be stochastic and can potentially be some known disturbance.


Turning to the GE for General Equilibrium, we note that for both models the equilibria are general in the sense that they both have separate financial and goods markets.  However, whereas the existence of an equilibrium state is \textit{assumed} in a DSGE model, in the network, it is an \textit{emergent} phenomenon, arising from the interactions among the agents.  Therefore, the existence of an equilibrium state is not a precondition for the functioning of a network and even chaotic systems or those with multistability can be analyzed.  


To model a sector, the DSGE model relies on the concept of a \textit{representative} agent.  The network model, instead, relies on a description in terms of two 
\textit{levels of abstraction}.  This allows us to understand and interpret the sector's macro-level behavior as a synthesis of the behavior at the lower levels of abstraction. 
In the network, the financial sector has two levels between macro and micro.  First, the entire sector is made up of a banking sector whose behavior is manifestly distinct from the securities market. Then, the banking sector consists of two different types: traditional and discount banks.   Similarly, the securities market consists of the stock and bond markets, with only the former exhibiting price volatility. If needed, a representative agent for the sector, or any subsectors for that matter, is then recovered by aggregating back up (see Section \ref{sec:InteractionsInterconnections}).   


\bibliography{refs} 
\bibliographystyle{ieeetr}

\end{document}